\documentclass[12pt]{article}
\usepackage[usenames,dvipsnames]{xcolor}
\usepackage{tikz}
\usetikzlibrary{matrix,arrows,positioning,shapes,snakes,fadings,decorations.pathmorphing,
decorations.pathreplacing,automata,shadows,decorations.markings,patterns,decorations.text,calc,angles,intersections,quotes,decorations.markings}
\usepackage{tkz-euclide}
\usepackage{pgfplots}
\usepackage{jheppub}
\usepackage{amsmath,amssymb,euscript,array,mathrsfs,appendix,ctable,mathabx}
\usepackage{mathtools,float}
\usepackage{arydshln}
\usepackage{todonotes}
\usepackage{graphicx}
\usepackage{wrapfig}
\usepackage{enumitem}
\usepackage{multirow}
\usepackage{bm,upgreek}
\usepackage{stmaryrd}
\usepackage{verbatim}

\usepackage{caption}
\usepackage{subcaption}

\usepackage{amsthm}

\makeatletter
\newsavebox{\@brx}
\newcommand{\llangle}[1][]{\savebox{\@brx}{\(\m@th{#1\langle}\)}%
  \mathopen{\copy\@brx\kern-0.6\wd\@brx\usebox{\@brx}}}
\newcommand{\rrangle}[1][]{\savebox{\@brx}{\(\m@th{#1\rangle}\)}%
  \mathclose{\copy\@brx\kern-0.6\wd\@brx\usebox{\@brx}}}
\makeatother

\newcommand \arXiv [1]{\href{http://arxiv.org/abs/#1}{\tt arXiv:#1}}

\usepackage{titlesec}
\titleformat{\subsection}[display]{\it}{}{0.1cm}{\vspace{-1.5cm}\begin{center}\thesubsection\hspace{0.2cm}}[\end{center}\vspace{-0.5cm}]

\newcommand{\PR}{ P}

\newcommand{\ket}[1]{|#1\rangle}
\newcommand{\bra}[1]{\langle#1|}
\newcommand{\tr}{\text{Tr}}

\newcommand\IM{\text{Im}\,}
\newcommand\RE{\text{Re}\,}
\newcommand\Bpi{{\boldsymbol\pi}}
\newcommand\Bsigma{{\boldsymbol\sigma}}
\newcommand\Be{{\boldsymbol e}}
\newcommand\Beta{{\boldsymbol\eta}}

\newcommand\SN{\text{sn}}

\newcommand{\EQ}[1]{\begin{equation}\begin{split} #1
\end{split}\end{equation}}

\title{Projections and teleportation of operator quenches in CFT}
\author{Zsolt Gyongyosi, Timothy J. Hollowood and S. Prem Kumar}
\affiliation{Department of Physics,\\ Swansea University,\\ Singleton Park,\\
Swansea, SA2 8PP, U.K.}
\emailAdd{z.gyongyosi.2133547@swansea.ac.uk,t.hollowood@swansea.ac.uk, s.p.kumar@swansea.ac.uk}
\abstract{Motivated by recent proposals for information recovery from black holes via non-isometric maps and post-selection in an effective description, we set up and investigate a teleportation scenario in a 2d CFT involving a local operator quench and  projection on a portion of space onto a Cardy state with the theory in the vacuum state. Using conformal invariance the system can be mapped to CFT with boundary (BCFT). R\'enyi entropies for spatial intervals in the projected state can then be computed as a function of the location of the quench,  either using the replica method, or using twist fields, the latter employing universal results for correlators at large $c$. We find qualitatively distinct behaviours in the two systems. Our replica computations reveal a surprising universal $n$ dependence of R\'enyi entropies which implies that  teleportation does occur but is not optimal as would be expected because the projector is not especially tuned. We also find that the curious $n$ dependence of the R\'enyi entropies means that the limit to the von Neumann entropy is not straightforward.
}

\notoc
\begin{document}

\maketitle

\newpage 

%\tableofcontents 

\section{Introduction}

\begin{figure}[ht]
\begin{center}
\begin{tikzpicture}[scale=2,every node/.style={scale=0.8},decoration={markings,mark=at position 0.5 with {\arrow{>}}}]
%\draw[very thick,black!20,->]  (-3,0) -- (3,0);
%\draw[very thick,black!20,->]  (0,-3) -- (0,3);
%
\draw[very thick,white,fill=yellow!10!white]  (-0.5,1.5) -- (1.5,1.5) -- (3,0) -- (2,-1) -- (-0.5,-1); 
%\draw[very thick,white,fill=yellow!10!white]  (0,-3) -- (0,1) -- (2,-1) -- (0,-1); 
%
\draw[very thick, dashdotted] (-0.5,-1) -- (-0.5,1.5);
\draw[very thick,black!40] (1.5,1.5) -- (3,0) -- (2,-1);
\draw[very thick,black!40,dashed] (2,-1) -- (-0.5,-1);
\draw[very thick,black!40] (1.5,1.5) -- (-0.5,-0.5);
\draw[very thick,decorate,
decoration={snake,amplitude=1pt, segment length=5pt}] (-0.5,1.5) -- (1.5,1.5);
\draw[red,thick,postaction={decorate}] (2.5,-0.5) -- (0.5,1.5);
\draw[blue,thick,postaction={decorate}] (-0.15,0.85) -- (0.5,1.5);
\draw[blue,thick,postaction={decorate}] (0.85,-0.15) -- (2,1);
\draw[blue,very thick,dashed] (-0.15,0.85) -- (0.85,-0.15);
\node at (2.4,1) {$\mathscr I^+$};
\node at (2.6,-0.7) {$\mathscr I^-$};
%
%\filldraw[black] (-0.5,1.5) circle (0.04cm);
\filldraw[black] (1.5,1.5) circle (0.04cm);
\node[rotate=45] at (0.8,0.65) {\footnotesize horizon};
\node[] at (0.75,1.65) {\footnotesize singularity};
\node[blue,rotate=-45] at (0.28,0.28) {\footnotesize entanglement};
\node[rotate=45,blue] at (0.9,0.1) {$A$};
\node[rotate=45,blue] at (0.3,1) {$\PR$};
\node[rotate=-45,red] at (2.1,0.1) {$\bar V$};
\end{tikzpicture}
\caption{\footnotesize The diagram shows an outgoing Hawking mode $A$ and its entangled partner $\PR$ falling into the singularity. There is an infalling system $\bar V$. In the final state projection scenario, a projection at the singularity onto a maximally mixed state of $\PR$ and $\bar V$ teleports the state of $\bar V$ to the outgoing Hawking radiation $A$.}
\label{fig1} 
\end{center}
\end{figure}
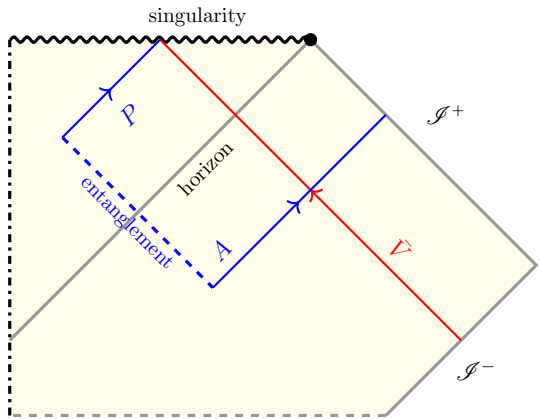

One approach to resolving the black hole information loss paradox is the so-called `final state proposal' of Horowitz and Maldacena \cite{Horowitz:2003he} (see also \cite{Lloyd:2013bza} for a thorough analysis). The idea is that at the level of the effective theory, on a Cauchy slice that includes what would be the classical singularity there is a projection on the quantum state along the singularity onto a maximally mixed state of the infalling modes $\bar V$ (left-moving on the Penrose diagram) and the infalling partners $\PR$ of the outgoing Hawking modes $A$ (right-moving on the Penrose diagram): see figure \ref{fig1}. Since the infalling partner modes $\PR$ and the outgoing Hawking modes $A$ are entangled, such a projection has the effect of teleporting the infalling quantum state to the outgoing Hawking radiation and so the `information' of the infalling modes is returned via the radiation. A rather puzzling feature is that in order for the teleportation to be 100\% efficient requires a very specific type of projector that is tuned to the entangled state of the $A$ and $P$ modes. Of course, such a projection violates the usual rules of quantum mechanics and leads to the instantaneous, acausal transfer of quantum information out of the black hole.\footnote{This is unlike the teleportation protocol in quantum mechanics which involves a measurement, and so a family of projections (in the simplest case) that can all be realized with certain probabilities. Since all possibilities are summed over, for a causally separated observer, there is no breakdown of casuality in this case. It is only after the measurement result has been revealed and then communicated is it meaningful to post select the state. For a single projection, the state of a causally separated observer can be instantaneously changed.} 

A refinement of the idea of final state projection, has recently been proposed in \cite{Akers:2022qdl}. This more recent work provides a reason why the effective theory would be subject to a projection even when the underlying microscopic description is perfectly consistent with unitarity and also naturally solves the puzzle of the specificity of the projector. In order to understand this, we need to think how the effective theory is related to the microscopic theory at the level of a Hilbert space description. One would expect that the Hilbert space of the effective theory ${\cal H}_\text{eff}$ is mapped into the Hilbert space of the microscopic theory ${\cal H}_\text{micro}$. Usually one would expect that this map, $V$, would be {\it isometric\/}, so preserving inner products. The intuition being that ${\cal H}_\text{micro}$ would be expected to be vastly bigger than ${\cal H}_\text{eff}$. It is this intuition that breaks down for an evaporating black hole. As the black hole evaporates ${\cal H}_\text{micro}$, with dimension order $e^{S_\text{BH}}$, is reducing while the dimension of ${\cal H}_\text{eff}$ is order $e^{S_\text{rad}}$, where $S_\text{rad}$ is the entropy of the radiation. It is precisely at the Page time that the dimension of the microscopic Hilbert space is actually smaller than the space needed to describe the semi-classical excitations on the black hole interior: something must give. The map $V$ cannot preserve the inner product of states and so is actually a {\it non-isometric\/} map. A non-isometric map can be realized as a unitary map followed by a projection. So once again we have a projection which can teleport information from the infalling state to the Hawking radiation. What is novel is that in this scenario the projections occur in an ongoing fashion, little-by-little, as the black hole evaporates rather than at the singularity. It is particularly noteworthy that, the fact that there is a perfectly unitary microscopic  description implies that at the effective level, teleportation occurs with prefect efficacy. 

The original final state hypothesis and its refinement in the effective theory motivate us to investigate projections in QFT and specifically their ability to teleport information between causally disconnected regions. Rather remarkably, as described initially in \cite{Rajabpour:2015uqa,Rajabpour:2015xkj} and then in \cite{Numasawa:2016emc,Antonini:2023aza,Akal:2021dqt}, certain kinds of projections on some local region can be described in  conformal field theories (CFTs) in $1+1$ dimensions by using conformal transformation to map the projection set-up to a boundary conformal field theory (BCFT) set-up which have been thoroughly explored (e.g.~see monograph \cite{BCFTbook}). In this set up, the projections are to a special set of states in the CFT known as Cardy states \cite{Cardy:1989ir}. These states are rather remarkable in that, unlike the vacuum state, they have no spatial entanglement, rather they have maximal entanglement between the chiral and anti-chiral sectors, or between the left- and right-moving sectors in Minkowski space \cite{Miyaji:2014mca}. In this sense, they are analogue of the final state projectors in the black hole scenario which are maximally entangled between the infalling modes $\bar V$ (so left-moving in figure \ref{fig1}) and the partners of the Hawking modes $\PR$ (so right-moving in figure \ref{fig1}).

In order to understand how projections can lead to the non-local flow of quantum information, we will consider their effect on an `operator quench' \cite{Asplund:2011cq, Nozaki:2013wia,Nozaki:2014hna,Calabrese:2016xau}. Before delving into the field theory scenario, let us describe how this works in the context of finite dimensional quantum system. In the model, there is a quantum system, which consists of two subsystems $A$ and $P$, in a pure entangled state $\ket{\psi}$ . There is an additional quantum system $\bar V$, the analogue of the infalling system in the state $\ket{\gamma}$. We then act with a projection $\mathscr P$ on $P\cup \bar V$, to model the action of the final state projection. So the state of the total system is
\EQ{
{\mathscr P}_{P\bar V}\ket{\psi}_{AP}\otimes\ket{\gamma}_{\bar V}\ .
}
Before the projection, the subsystem $A$ has no information regarding the state $\ket{\gamma}$ but after the projection the information needed to specify the state can be `teleported' to $A$ depending on the state $\ket{\psi}$ and the projection $\mathscr P$. An efficient way to compute whether an arbitrary state $\ket{\gamma}$ can be reconstructed from the reduced state on $A$, $\rho_A$, is to maximally entangle $\bar V$ with an auxiliary system $V$ of the same dimension. Before the projection, the mutual information $I(V,A)=0$ because the subsystem $A$ has no correlation with $\bar V$. After the projection, we can say that teleportation is completely successful if $I(V,A)=2S(V)$ which means that, whatever the state of $\bar V$ is, it can be reconstructed from $A$. We are assuming here that $\bar V$ (and hence $V$) are small systems compared with $A$ and $P$. Given that $I(V,A)=S(V)+S(A)-S(V\cup A)$, and given that $S(V)$ is fixed and unaffected by the projection, a measure of teleportation involves the entropy difference
\EQ{
\Delta S\equiv S(V\cup A)-S(A)\ .
\label{cut}
} 
Before projection this is equal to $S(V)$ and after projection, perfect teleportation requires this to jump to $-S(V)$.

In order to set up the scenario in a 2d CFT, the analogue of the state $\ket{\psi}$, in the first instance, will be chosen to be the vacuum state with $A$ and $P$ associated to a distinct spatial subsets on a Cauchy surface. The idea is that $A$ is modelling some region outside the black hole and $P$ the inside of the black hole.\footnote{In our scenario, there is a third sub-region, the gap between $P$ and $A$.} In the CFT the infalling subsystem $\bar V$ will be modelled by acting on the vacuum state by a set of left-moving (anti-chiral) operators $\bar V_a(\bar w)\ket{0}$ locally at spatial position $\bar w$ (and taken at a time $t=0$).\footnote{There will need to be a small imaginary shift in the position in order that the state is normalizable as we describe in following sections.} The auxiliary system $V$ will then be associated to a set of right-moving (chiral) operators acting on the vacuum $V_a(w)\ket{0}$ at spatial point $w$. The entanglement between $V$ and $\bar V$ can be established by defining an operator and state
\EQ{
\ket{{\cal O}}\equiv{\cal O}(w,\bar w)\ket{0}={\cal N}\sum_{a=1}^{d_{\cal O}}V_a(w)\bar V_a(\bar w)\ket{0}\ ,
\label{sta2}
}
where $V_a$ and $\bar V_a$ are suitably normalized and ${\cal N}$ is an overall normalization factor computed in later sections. If the operator is inserted at some initial time $<0$, at $t=0$ the chiral and anti-chiral components 
are localized at different points on the Cauchy slice and we can use this freedom to mimic the teleportation scenario. The quantity $d_{\cal O}$ is known as the quantum dimension of the operator and the state $\ket{{\cal O}}$ has entanglement entropy between the chiral and anti-chiral sectors that we identify as $S(V)=\log d_{\cal O}$. 

Before we consider teleportation by making a projection, we can compute entropies of the subsystem on the state involving the operator quench \eqref{sta2}:
\EQ{
S^{\ket{{\cal O}}}(A,w,\bar w)\equiv S[\tr_{A^c}\ket{{\cal O}}\bra{{\cal O}}]\ ,
\label{kiw}
}
where the von Neumann entropy $S[\rho]=-\tr \rho\log\rho$. This entropy has the usual UV divergence of the entropy of a spatial region of a field theory which does not depend on the operator quench and can be cancelled by subtracting the entropy of $A$ in the vacuum state, what we call the entropy excess of the quench,
\EQ{
S^{\ket{{\cal O}}}(A,w,\bar w)_\text{reg}\equiv S[\tr_{A^c}\ket{{\cal O}}\bra{{\cal O}}]-S[\tr_{A^c}\ket{0}\bra{0}]\ .
\label{kiw2}
}
However, the effect of the operator quench can be deduced from computing entropy differences that are UV safe. In particular, the entropy difference
\EQ{
\Delta S^{\ket{{\cal O}}}\equiv S^{\ket{{\cal O}}}(A,w\in A,\bar w\in A^c)-S^{\ket{{\cal O}}}(A,w\in A^c,\bar w\in A^c)\ ,
\label{fet}
}
is such a UV safe quantity. We will review the analysis of \cite{Nozaki:2014hna,He:2014mwa} in section \ref{s2} and show how to compute this difference. One finds that $\Delta S^{\ket{{\cal O}}}=\log d_{\cal O}$ which shows that, as expected, $V$ is not correlated with $A$, the mutual information $I(\bar V,A)=S(V)-\Delta S^{\ket{{\cal O}}}=0$.

Now we make the projection $\mathscr P$ at $t=0$ on a disjoint region $P$ to $A$, modelling the inside of the black hole. The state of the CFT is now
\EQ{
\ket{{\mathscr P}{\cal O}}\equiv{\mathscr P}{\cal O}(w,\bar w)\ket{0}\ .
\label{we2r}
}
and now the entropy \eqref{kiw} is
\EQ{
S^{\ket{{\mathscr P}{\cal O}}}(A,w,\bar w)\equiv S[\tr_{A^c}\ket{{\mathscr P}{\cal O}}\bra{{\mathscr P}{\cal O}}]\ .
}
In order to mimic the infalling system $\bar V$, we choose the position of the anti-chiral operators in \eqref{sta2} to be $\bar w\in P$. Then by choosing the chiral insertion at either $w\in A$ or $w\notin A\cup P$, the CFT entropy of $A$ is the analogue of $S(V\cup A)$ and $S(A)$, respectively. This motivates us, as in \eqref{fet}, to define the UV finite entropy difference
\EQ{
\Delta S^{\ket{{\mathscr P}{\cal O}}}(\bar w)=S^{\ket{{\mathscr P}{\cal O}}}(A,w\in A,\bar w)-S^{\ket{{\mathscr P}{\cal O}}}(A,w\notin A\cup P,\bar w)\ ,
\label{fat}
}
as a measure of the success of teleportation. Note that being an entropy difference means that it is a UV safe quantity and we do not have to use the subtracted entropies. The entropy difference will depend on exactly where the anti-chiral operators are inserted in $P$ and perfect teleportation would require, as in the finite-dimensional example, that $\Delta S^{\ket{{\mathscr P}{\cal O}}}(\bar w\in P)=-\log d_{\cal O}$. We can also take $\bar w$ outside $P$ (and not in $A$) and then we would expect the entropy difference to be $\Delta S^{\ket{{\mathscr P}{\cal O}}}(\bar w\notin P\cup A)=\log d_{\cal O}$. In section \ref{s3}, we will compute the R\'enyi entropy version of $\Delta S^{\ket{{\mathscr P}{\cal O}}}$ as a function of $\bar w$ starting at the edge of the region $P$ and then moving inside it. Success of teleportation would require this to jump from $\log d_{\cal O}$ to $-\log d_{\cal O}$.

The conventional approach to computing an entropy like $S(A)$ in a QFT would be to compute the R\'enyi entropy $S_n(A)$, using the replica method, and then take an analytic continuation $n\to1$ relying on Carlson's Theorem. We will find, within the free scalar field theory setting, that for the R\'enyi entropy generalization of the entropy difference \eqref{fat}, the dependence on $n$ is not simple and a straighforward analytic continuation to $n=1$ and von Neumann entropy is ambiguous.  For large-$c$ sparse CFTs,
we will employ knowledge of correlators  involving twist fields (with integer $n$) and semiclassical conformal blocks to infer the behaviour of R\'enyi entropies in the projected state, and find qualitatively distinct features relative to the free scalar theory.

The paper is organized as follows: in section 2 we review, based on the work of \cite{Nozaki:2014hna,He:2014mwa}, the computation of the R\'enyi entropy of an operator quench with a scenario without a projection. In section 3, we add in the projection in a semi-infinite spatial region and where the entropy region is also semi-infinite. In section 4, we take a different approach and compute the R\'enyi entropies using twist operators in a large $c$ CFT where universal statements can be made. In section 5, we consider the case with finite intervals but only for the second R\'enyi entropy where a uniformization map can be written in terms of elliptic functions. Finally in section 6 we draw some conclusions.

\section{Entropy of an operator quench}\label{s2}

In this section, we review relevant aspects of the entropy excesses associated to operator quenches, but without projectors, as computed in \cite{Nozaki:2014hna,He:2014mwa}. The context is $1+1$-dimensional CFT. We use the chiral/anti-chiral coordinates which in Minkowski space are associated to the right- and left-moving sectors, of the theory, $w=-t+x$ and $\bar w=t+x$. In particular, in Lorentzian signature $\bar w$ does not denote complex conjugation. Consider the state consisting of an operator acting on the vacuum state
\EQ{
\ket{{\cal O}}={\cal O}(w_1,\bar w_1)\ket{0}={\cal N}\sum_{a=1}^{d_{\cal O}}V_a(w_1)\bar V_a(\bar w_1)\ket{0}\ .
\label{sta}
}
The sum here is over $d_{\cal O}$ component chiral/anti-chiral insertions and this is the {\it quantum dimension\/} of the state. The constant ${\cal N}$ is a normalization factor.  With the normalization of the operators as in \eqref{ope} below, the resulting state is a maximally entangled state between the chiral and anti-chiral sectors with R\'enyi entropy $S_n(V)=\log d_{\cal O}$. 

In order to make the state normalizable, one introduces a small regulator $\epsilon$ which sets the length scale of the energy density support of the disturbance caused by the operator; so for real $w$ and $\bar w$,
\EQ{
w_1=w+i\epsilon\ ,\qquad \bar w_1=\bar w-i\epsilon\ .
}
Note that the anti-chiral (left-moving) component is localized around $x=\bar w-t$ while the chiral (right-moving) component is localized around $x=w+t$. So we can interpret the state as being due to a local operator insertion at time $t=\frac12(\bar w-w)$ at position $x=\frac12(w+\bar w)$. As time evolves the chiral and anti-components then separate. It is the entanglement between the chiral and anti-chiral components that are separated in space that will be exploited. 

The associated bra state is 
\EQ{
\bra{0}{\cal O}^\dagger(w_2,\bar w_2)=\Big({\cal O}(w_1,\bar w_1)\ket{0}\Big)^\dagger\ ,
}
where
\EQ{
w_2=w_1^*=w-i\epsilon\ ,\qquad\bar w_2=\bar w_1^*=\bar w+i\epsilon\ .
\label{cjn}
}
Note that complex conjugation is indicated as $w^*$ and not $\bar w$. The normalization factor can be computed from the OPEs 
\EQ{
V_a^\dagger(w_2)V_b(w_1)=(-2i\epsilon)^{-2\Delta}\delta_{ab}\ ,\qquad \bar V^\dagger_a(\bar w_2)\bar V_b(\bar w_1)=(2i\epsilon)^{-2\Delta}\delta_{ab}\ ,
\label{ope}
}
yielding
\EQ{
{\cal N}=\frac{(2\epsilon)^{2\Delta}}{\sqrt{d_{\cal O}}}\ .
}

\subsection{An example: free scalar CFT}\label{s2.1}

It is useful to have a tractable example in mind for the analysis. To this end, we take, as in \cite{He:2014mwa}, a massless (uncompactified) scalar field. A convenient choice of operators which have anti-chiral/chiral entanglement are 
\EQ{
{\cal O}={\cal N}\big(e^{i\alpha\phi}+e^{-i\alpha\phi}\big)\ .
}
In this case, the conformal dimensions are $\Delta=\bar\Delta=\frac{\alpha^2}2$ and the quantum dimension is $d_{\cal O}=2$. What makes this choice of CFT and operator appealing, is that we have explicit expressions for the correlators on the complex $z$ plane
\EQ{
\big\langle\prod_{j=1}^{n}e^{i\alpha_j\phi(z_j,\bar z_j)}\big\rangle_{{\mathbb C}}=\prod_{j<k=1}^{n}|z_{jk}|^{2\alpha_j\alpha_k}\delta\big(\sum_j\alpha_j\big)\ ,
\label{gug}
}
from which we have
\EQ{
\big\langle\prod_{j=1}^{n}{\cal O}(z_j,\bar z_j)\big\rangle_{{\mathbb C}}={\cal N}^{n}\sum_{e_j=\pm1\atop \sum e_j=0}\prod_{j<k=1}^{n}|z_{jk}|^{2\alpha^2e_je_k}\ ,
\label{gug}
}
where $z_{jk}=z_j-z_k$. Later the $z$ coordinate will be identified with that of a uniformized geometry.

\subsection{Entropy of a local quench}

The entanglement between the chiral and anti-chiral sectors can be measured by computing entropies of sub-regions that may contain $w$, $\bar w$ or both. Of course these entropies will suffer from UV divergences. However, one can define a UV safe quantity which essentially subtracts the vacuum contribution to the entropy to leave the excess entropy that is caused by the operator quench. The R\'enyi excess entropy of a spatial region $A$ in the presence of the operator quench, is formally defined as
\EQ{
S^{\ket{{\cal O}}}_n(A,w,\bar w)_\text{reg}=S_n\big[\tr_{A^c}\ket{{\cal O}}\bra{{\cal O}}\big]-S_n[\tr_{A^c}\ket{0}\bra{0}\big]\ ,
\label{pep}
}
where $A^c$ is the complement of $A$ and $e^{(1-n)S_n[\rho]}=\tr\,\rho^n$. Of course, individually these quantities can only be properly defined in the theory with a suitable UV cut off, however, the cut-off dependent terms vanish for the difference to leave a finite answer.

The formalism proceeds via the replica method. If we wish to compute the R\'enyi entropy of a region $A$, then we consider $n$-fold cover $\Sigma_n$ of the original spacetime $\Sigma_1$ branched at the endpoints $\partial A$. This means that there will be $n$ copies of the insertion of the operator to create the replicated ket state and $n$ copies of the conjugate to create the bra. Let us denote the replicas of the pair $(w_1,\bar w_1)$ as $(w_{2j-1},\bar w_{2j-1})$ and of $(w_2,\bar w_2)$ as $(w_{2j},\bar w_{2j})$, $j=1,2,\ldots,n$, so given \eqref{cjn}
\EQ{
w_{2j}=w_{2j-1}^*\ ,\qquad \bar w_{2j}=\bar w^*_{2j-1}\ .
}
The R\'enyi excess entropy in the presence of the quench is then determined by the $2n$-point correlator on $\Sigma_n$:
\EQ{
S^{\ket{\cal O}}_n(A,w,\bar w)_\text{reg}=\frac1{1-n}\log\,\big\langle\prod_{j=1}^n{\cal O}(w_{2j-1},\bar w_{2j-1}){\cal O}^\dagger(w_{2j},\bar w_{2j})\big\rangle_{\Sigma_n}\ .
\label{cor}
}
Note that the subtraction of the vacuum entropy in \eqref{pep} is implicit in the standard normalization of the correlator relative to the vacuum.

Determining the quench entropy is now delightfully simple in the limit where the regulator $\epsilon\to0$. There are 4 cases to consider with either of $\bar w,w$ being in $A$ or not. Consider first of all, the case where the chiral  point $w\not\in A$. Then in the replica geometry $\Sigma_n$, $w_{2j-1}$ is close to $w_{2j}$, to order $\epsilon$. Hence, the chiral  sector is dominated by the bootstrap channel that pairs up points as $(12)(34)\cdots(2n-1,2n)$. We may invoke the OPE \eqref{ope} 
\EQ{
V^\dagger_{a_{2j}}(w_{2j})V_{a_{2j-1}}(w_{2j-1})=(-2i\epsilon)^{-2\Delta}\delta_{a_{2j-1}a_{2j}}\ .
\label{rep}
} 
In this case, the correlator of the chiral operators produces the product of delta functions $\prod_j\delta_{a_{2j-1}a_{2j}}$. 

On the other hand, if $w\in A$ then $w_{2j-1}$ and $w_{2j}$ are separated by the cut along $A$ and now $w_{2j}$ is close to the replica point $w_{2j+1}$, rather than $w_{2j-1}$. Hence, the relevant OPE now is
\EQ{
V^\dagger_{a_{2j}}(w_{2j})V_{a_{2j+1}}(w_{2j+1})=(2i\epsilon)^{-2\Delta}\delta_{a_{2j+1}a_{2j}}\ .
} 
(The indices $j$ are understood mod $n$). So in this case the channel that dominates pairs up the points as  $(1,2n)(32)\cdots(2n-1,2n-2)$. In this case, the correlator of the chiral operators produces the product of delta functions $\prod_j\delta_{a_{2j+1}a_{2j}}$. 

The same analysis holds for the anti-chiral sector. In the final analysis, all the $(\pm2i\epsilon)^{-2\Delta}$ factor are just cancelled by the normalization ${\cal N}$ factors, so what remains are the products of delta functions.

\subsection{Bootstrap channels and permutations}\label{s2.3}

Let us pause, to set up a general way to describe bootstrap channels more systematically in the present context. A general one corresponds to a limit where the odd points asymptote to the even points in pairs $w_{2j-1}\to w_{2k}$ and so the correlators are dominated by the OPEs
\EQ{
V_{2j-1}(w_{2j-1})V^\dagger_{2k}(w_{2k})\longrightarrow \frac1{w_{2j-1,2k}^{2\Delta}}\ .
}
The channel can be labelled by an element of the permutation group $\Bpi:\ j\to k$ on $n$ objects. The same can be said of the anti-chiral sector and so the chiral and anti-chiral sectors have their own channel which we denote by a pair of elements of the symmetric group $\Bpi$ and $ \Bsigma$, respectively.

For a general channel $(\Bpi, \Bsigma)$, all the operators are paired up in a way dictated by these elements and we may invoke the OPE \eqref{ope} 
\EQ{
&V^\dagger_{a_{2\Bpi(j)}}(w_{2\Bpi(j)})V_{a_{2j-1}}(w_{2j-1})=(-2i\epsilon)^{-2\Delta}\delta_{a_{2j-1},a_{2\Bpi(j)}}\ ,\\[5pt]
&\bar V^\dagger_{a_{2\Bsigma(j)}}(\bar w_{2\Bsigma(j)})\bar V_{a_{2j-1}}(\bar w_{2j-1})=(2i\epsilon)^{-2\Delta}\delta_{a_{2j-1},a_{2 \Bsigma(j)}}\ .
\label{rep2}
}
What results is a product of delta functions from the chiral and anti-chiral sectors that depend on two elements $\Bpi, \Bsigma\in \text{S}_n$. Finally we have
\EQ{
S_n^{\ket{\cal O}}(A,w,\bar w)_\text{reg}&=\frac1{1-n}\log \frac1{d_{\cal O}^n}\sum_{\{a_j\}}\prod_{j=1}^n\delta_{a_{2j-1},a_{2\Bpi(j)}}\delta_{a_{2j-1},a_{2 \Bsigma(j)}}\\[5pt]&=\frac1{1-n}\log \frac1{d_{\cal O}^n}\sum_{\{a_{2j}\}}\prod_{j=1}^n\delta_{a_{2\Bpi(j)},a_{2 \Bsigma(j)}}=\frac{{\mathscr D}(\Bpi, \Bsigma)}{n-1}\log d_{\cal O}\ ,
\label{get}
}
where ${\mathscr D}(\Bpi, \Bsigma)$ is the Cayley distance between the permutations, the minimum number of pairwise swaps needed to relate the two elements. This is $n$ minus the number of cycles of the element $\Bpi \Bsigma^{-1}$.

In the present case, there are two channels of interest. The first involves the limit $z_{2j-1}\to z_{2j}$ corresponding to the identity element $\Be\in\text{S}_n$. The other channel involves $z_{2j+1}\to z_{2j}$ corresponding to the cyclic permutation $\Beta\in\text{S}_n$: $n\to n-1\to\cdots\to 2\to 1\to n$. 
In present circumstances, the permutations $\Bpi$ and $ \Bsigma$ are chosen according to whether $w$ and $\bar w$ are in $A$ or not. Specifically, if $w\not\in A$, $\Bpi=\Be$, while if $w\in A$ then $\Bpi=\Beta$. The same applies to the anti-chiral sector. Since ${\mathscr D}(\Be,\Be)={\mathscr D}(\Beta,\Beta)=0$ and  ${\mathscr D}(\Be,\Beta)={\mathscr D}(\Beta,\Be)=n-1$, we have 
\EQ{
S_n^{\ket{\cal O}}(A,w,\bar w)_\text{reg}=\begin{cases} 0 &w,\bar w\in A\quad\text{or}\quad w,\bar w\not\in A\ ,\\
\log d_{\cal O} & w\in A,\bar w\not\in A\ ,\quad\text{or vice-versa}\ .\end{cases}
\label{kep}
}
The interpretation is that the UV safe quantity, the R\'enyi excess, $\delta S^{\ket{\cal O}}_n(A)$ picks up the entanglement entropy between the chiral and anti-chiral sectors when $A$ contains one of the insertions, chiral or anti-chiral, but not both. It is worth remarking that entropy differences, for example 
\EQ{
\Delta S_n^{\ket{\cal O}}(\bar w)=S^{\ket{\cal O}}_n(A,w\in A,\bar w)-S^{\ket{\cal O}}(A,w\in A^c,\bar w)\ ,
}
which equals $\log d_{\cal O}$ if $\bar w\in A^c$ and $-\log d_{\cal O}$ if $\bar w\in A$, is a UV safe quantity for which the vacuum subtractions cancel out. This quantity, with $\bar w\in A^c$, is precisely the analogue of the entropy difference \eqref{cut} in the finite dimensional example (before projection).

\section{Including a projection and teleportation}\label{s3}

Now we consider the effect of a projector. The projector is onto a Cardy state $\ket{B}$ made in the region which we take to be $P=[-\infty,0]$. 
\begin{figure}[ht]
\begin{center}
\begin{tikzpicture} [scale=0.9,every node/.style={scale=0.8}]
\draw[fill = Plum!10!white, draw = Plum!10!white] (-2,-3) rectangle (8,3);
\draw[fill = white, draw = white] (-2,-0.1) rectangle (1,0.1);
\draw[black!40,thick] (1,0) --(8,0);
\draw[black!40,thick] (1,-3) -- (1,3);
\draw[line width=0.6mm,blue] (5,0) -- (8,0);
\draw[line width=0.6mm,red] (-2,0.1) -- (1,0.1);
\draw[line width=0.6mm,green] (-2,-0.1) -- (1,-0.1);
\node at (5,0.3) {$a$};
\filldraw[black] (6,0.2) circle (2pt);
\filldraw[black] (6,-0.2) circle (2pt);
\filldraw[black] (-0.5,0.2) circle (2pt);
\filldraw[black] (-0.5,-0.2) circle (2pt);
\filldraw[black] (1.2,0.2) circle (2pt);
\filldraw[black] (1.2,-0.2) circle (2pt);
\draw[->,thick,densely dashed] (1.2,0.2) to[out=180,in=0] (-0.4,0.2);
\draw[->,thick,densely dashed] (1.2,-0.2) to[out=180,in=0] (-0.4,-0.2);
\filldraw[purple] (4,0.15) circle (2pt);
\filldraw[purple] (4,-0.15) circle (2pt);
\node at (6,0.5) {$w_1$};
\node at  (6,-0.5) {$w_2$};
\node at (4,0.5) {$w_1$};
\node at  (4,-0.5) {$w_2$};
\node at (0.2,0.5) {$\bar w_2$};
\node at  (0.2,-0.5) {$\bar w_1$};
\node at (-1,0.5) {$\PR$};
\node at (7.2,0.3) {$A$};
\end{tikzpicture}
\caption{\footnotesize The scenario of interest. There is a projection slit $\PR=[-\infty,0]$ shown in red/green. The chiral  points $w_1=w+i\epsilon$ and $w_2=w-i\epsilon$ are either in the entropy region $A$ (black) or not (purple). We will consider how things vary with the position of the anti-chiral points $\bar w_1=\bar w-i\epsilon$ and $\bar w_2=\bar w+i\epsilon$ which lie on either side of the projection slit, as shown.}
\label{fig3} 
\end{center}
\end{figure}
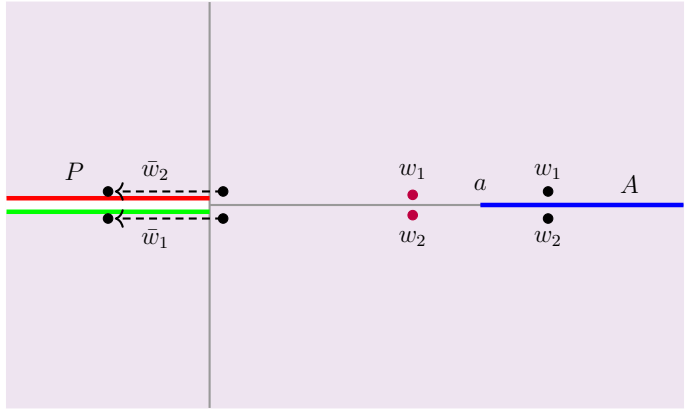

The way to introduce the projector onto a Cardy state \cite{Cardy:1989ir} in a CFT was described in \cite{Rajabpour:2015uqa,Rajabpour:2015xkj,Numasawa:2016emc}. The idea is to cut open along $\PR$ to create a slit and then impose appropriate boundary conditions along the top and the bottom of the slit that corresponds to inserting $\ket{B}\bra{B}$ along $\PR$. In the presence of the operator quench, the state is now
\EQ{
\ket{{\mathscr P}{\cal O}}={\mathscr P}{\cal O}(w_1,\bar w_1)\ket{0}\ ,
\label{wer}
}
where the projection
\EQ{
{\mathscr P}=\ket{B}\bra{B}_{\PR}\otimes I_{\PR^c}\ .
}
The state \eqref{wer} describes a local operator insertion inserted at $x=\frac12(w+\bar w)$ at time $t=\frac12(\bar w-w)$ which is then projected at $t=0$. In this case, unlike the case without a projection, the $t$ dependence of the state is not literally the time evolution of the state because the projector does not commute with the Hamiltonian. 
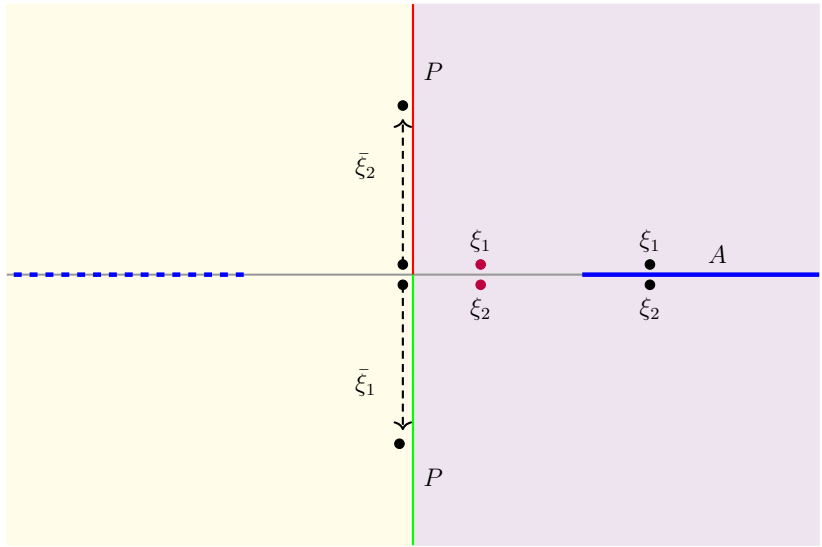
\begin{figure}[ht]
\begin{center}
\begin{tikzpicture} [scale=0.9,every node/.style={scale=0.8}]
\draw[fill = Plum!10!white, draw = Plum!10!white] (0,-4) rectangle (6,4);
\draw[fill = yellow!10!white, draw = yellow!10!white] (0,-4) rectangle (-6,4);
\draw[black!40,thick] (-6,0) --(6,0);
\draw[line width=0.6mm,red,thick] (0,0) -- (0,4);
\draw[line width=0.6mm,green,thick] (0,-4) -- (0,0);
\draw[line width=0.6mm,blue] (2.5,0) -- (6,0);
\draw[line width=0.6mm,blue,dashed] (-2.5,0) -- (-6,0);
%\node at (2.5,0.5) {$\sqrt{a}$};
%\node at (-2,0.5) {$-\sqrt a$};
%
%\begin{scope}[xshift=-2cm]
%\node at (6.5,3.5) {$\xi$};
%\draw[-] (6.2,3.9) -- (6.2,3.2) -- (6.8,3.2);
%\end{scope}
%
\filldraw[black] (3.5,0.15) circle (2pt);
\filldraw[black] (3.5,-0.15) circle (2pt);
\filldraw[black] (-0.15,2.5) circle (2pt);
\filldraw[black] (-0.2,-2.5) circle (2pt);
\filldraw[black] (-0.15,0.15) circle (2pt);
\filldraw[black] (-0.15,-0.15) circle (2pt);
\filldraw[purple] (1,0.15) circle (2pt);
\filldraw[purple] (1,-0.15) circle (2pt);
\draw[->,thick,densely dashed] (-0.15,0.15) -- (-0.15,2.3);
\draw[->,thick,densely dashed] (-0.15,-0.15) -- (-0.15,-2.3);
\node at (3.5,0.5) {$\xi_1$};
\node at  (3.5,-0.5) {$\xi_2$};
\node at (1,0.5) {$\xi_1$};
\node at  (1,-0.5) {$\xi_2$};
\node at (-0.7,1.6) {$\bar\xi_2$};
\node at  (-0.7,-1.6) {$\bar\xi_1$};
%\node at (5.5,0) {$C$};
\node at (0.3,3) {$\PR$};
\node at (0.3,-3) {$\PR$};
%\node at (0.75,0) {$B$};
\node at (4.5,0.3) {$A$};
%\node[blue] at (4.5,-0.4) {$A_-$};
%
%\node at (-3,3.5) {image};
\end{tikzpicture}
\caption{\footnotesize The map to the BCFT interpretation in the $\xi$ plane which has opened up the projection slit to the  imaginary axis. When $\bar w\in P$, the anti-chiral points lie on the imaginary axis. Here we show the images of the antichiral points, interpreted as chiral coordinates in the left half plane, employing  the doubling trick for BCFT.  The dashed blue line is the image of the interval $A$ in the doubled picture.}
\label{fig4} 
\end{center}
\end{figure}

The significance of choosing Cardy states \cite{Cardy:1989ir} is that these preserve some amount of the conformal symmetry and we can conformally map the $z$ plane with slit to the region $\RE\xi\geq0$ of the $\xi$ plane by the 
conformal map
\EQ{
w=\xi^2\ .\label{openup}
}
The upper/lower boundaries of the slit are mapped to the positive/negative imaginary axis, a shown in figure \ref{fig4}. \
This leads to formulation of the CFT in the presence of a boundary within the  well understood boundary CFT (BCFT) framework \cite{Cardy:1989ir,Cardy:1984bb,Cardy:1986gw,Cardy:2004hm}.

In order to compute the R\'enyi excess entropies we have to evaluate correlators in the appropriately replicated BCFT. An important ingredient in this evaluation  is the so-called `doubling trick' in which the anti-chiral operators $\bar V_a(\bar\xi)$ are mapped to their image points, 
\EQ
{
\RE\,\bar\xi\to -\RE\,\bar\xi\,,
}
upon reflection across the boundary at $\RE\xi=0$, and a general correlator in BCFT is then  viewed as a chiral CFT correlator on the plane \cite{BCFTbook, Sully:2020pza}. Thus chiral and anti-chiral operators can now have a non-trivial OPE.\footnote{Note, $\bar\xi$, like $\bar w$, is just a left-moving null coordinate and not the complex conjugate of $\xi$.} However, if we choose the chiral points to be outside the projection region $w\not\in\PR$ they will always be in close pairs, as shown in figures \ref{fig3} and \ref{fig4} in the $z$ and $\xi$ planes, respectively. The chiral operators are then in a definite channel, depending upon whether $w\in A$, or $\not\in A$, and, therefore it is never necessary to consider non-trivial OPEs between the chiral and anti-chiral operators. In addition, we need only consider the case where the anti-chiral points lie on the boundary itself, parametrised by $\gamma \in [0,\frac\pi2]$\,, 
\EQ{
\bar\xi_{2j-1}=-i\sqrt a\tan\gamma\ ,\qquad \bar\xi_{2j}=i\sqrt a\tan\gamma\,.
\label{taj}
}
Note that because these points are separated in the $\xi$ plane we do not need to include the small shifts $\pm i\epsilon$ in the $w$ plane. This parametrisation naturally follows from mapping the half-plane to the interior of the unit circle and  $\gamma$ is then the polar angle along the unit circle. The limit $\gamma=0$ has the anti-chiral insertion on the edge of $\PR$ and as the angle $\gamma$ increases the points move further into $P$. Note that 
\EQ{
\tan\gamma=\sqrt{\frac{|\bar w|}a}\ .
}
for $\bar w\in[-\infty,0]$.

\subsection{The $n=2$ case in scalar CFT}

As a warm up, in this section, we consider the case of the second R\'enyi entropy in the scalar CFT we introduced in section \ref{s2.1}. We will compute the R\'enyi excess entropies for $w\in A$ and $w\not\in A$ separately and then compute the difference \eqref{fat}. In order to compute the correlators, we need to map the replica geometry, the two copies of the $\xi$-plane branched at $\xi=\pm\sqrt a$ (image of $w=a$ and its doubled copy) to a single copy of the complex $z$ plane using a suitable map, e.g.
\EQ{
z^2=\frac{\xi-\sqrt a}{\xi+\sqrt a}\ .
\label{nip}
}
On the $z$ plane, the boundary---the image of the original slit---is now the unit circle $|z|=1$. The disc $|z|\leq 1$ contains  the replicas of the original $w$ plane, i.e.~$\RE\xi>0$, and the complement of the disc is the image region of BCFT, $\RE\xi<0$. 

To begin, let us compute the normalization factors of the operators. These are computed in the un-replicated geometry $\Sigma_1$, the $\xi$ plane. In such a correlation function above, we must be careful: in the doubled formalism, we cannot in general use the result \eqref{gug} because now there is a non-trivial OPE between the chiral and anti-chiral operators $V_a$ and $\bar V_b$. However, we only need the correlator when the chiral points are in infinitesimal close pairs, and in this case the OPE between chiral and anti-chiral operators is never needed. Consequently, we can use \eqref{gug} to compute the correlators, and we find, 
\EQ{
\big\langle{\cal O}(\xi+\delta\xi,-i\sqrt a\tan\gamma){\cal O}(\xi-\delta\xi,i\sqrt a\tan\gamma)\big\rangle_{\Sigma_1}=2{\cal N}^2(-2\delta\xi\cdot 2i\sqrt a\tan\gamma)^{-\alpha^2}\,,
\label{er1}
}
which must be set to unity, fixing the value of the normalisation ${\cal N}$. Here $\xi=\sqrt w$ and $\delta\xi=i\epsilon/(2\sqrt w)$.

Under the uniformation map \eqref{nip}, the anti-chiral points \eqref{taj} map to 
\EQ{
\bar z_1=ie^{i\gamma}\ ,\quad\bar z_2=ie^{-i\gamma}\ ,\quad\bar z_3=-ie^{i\gamma}\ ,\quad\bar z_4=-ie^{-i\gamma}\ .
}
For the chiral points, there are two cases to consider depending upon whether the chiral point is (I) not in $A$, so $w\in[0,a]$, or (II) in $A$, so $w>a$. Let us take the former case (I) first. In this case, under the uniformization map \eqref{nip}, the chiral points are at
\EQ{
z_1=i\lambda+\delta z_1\ ,\quad z_2=i\lambda+\delta z_2\ ,\quad z_3=-i\lambda+\delta z_3\ ,\quad z_4=-i\lambda+\delta z_4\ ,
}
where
\EQ{
\lambda=\left|\frac{\sqrt w-\sqrt a}{\sqrt a+\sqrt w}\right|^{1/2}\ ,\qquad \delta z_j=\frac{dz_j}{d\xi_j}\,\delta\xi_j\ .
\label{pun}
}
and $\delta\xi_j=(\delta\xi,-\delta\xi,\delta\xi,-\delta\xi)$.

\begin{figure}[ht]
\begin{center}
\begin{tikzpicture} [scale=0.9,every node/.style={scale=0.8}]
\draw[fill=blue!10,blue!20,opacity=0.5] (2,1.5) -- (4.5,1.5) -- (4.5,-3.5) -- (2,-3.5) -- cycle;
\draw[fill=blue!10,blue!20,opacity=0.5] (10,1.5) -- (7.5,1.5) -- (7.5,-3.5) -- (10,-3.5) -- cycle;
\draw[fill = Plum!10!white, draw = Plum!10!white,opacity=0.5] (4,0) -- (8,0) -- (12,3) -- (8,3) -- cycle;
\draw[fill = yellow!10!white, draw = yellow!10!white,opacity=0.5] (0,0) -- (4,0) -- (8,3) -- (4,3) -- cycle;
\draw[black!30] (0,0) -- (8,0) -- (12,3) -- (4,3) -- cycle;
\draw[black!60] (2,1.5) -- (10,1.5);
\draw[thick,green] (4,0) -- (6,1.5);
\draw[thick,red] (6,1.5) -- (8,3);
\draw[thick,blue] (7.5,1.5) -- (10,1.5);
\draw[thick,blue,dashed] (2,1.5) -- (4.5,1.5);
\filldraw[black] (8.63,1.6) circle (2pt);
\filldraw[black] (8.37,1.4) circle (2pt);
\node at (8.63,1.9) {$\xi_1$};
\node at  (8.37,1.1) {$\xi_4$};
\filldraw[black] (7.2,2.4) circle (2pt);
\filldraw[black] (4.8,0.6) circle (2pt);
\node at (6.8,2.5) {$\bar\xi_2$};
\node at  (5.2,0.6) {$\bar\xi_1$};
\filldraw[purple] (6.93,1.6) circle (2pt);
\filldraw[purple] (6.67,1.4) circle (2pt);
\node at (6.93,1.9) {$\xi_1$};
\node at  (6.67,1.1) {$\xi_2$};
\begin{scope}[yshift=-5cm]
\draw[fill = Plum!10!white, draw = Plum!10!white,opacity=0.5] (4,0) -- (8,0) -- (12,3) -- (8,3) -- cycle;
\draw[fill = yellow!10!white, draw = yellow!10!white,opacity=0.5] (0,0) -- (4,0) -- (8,3) -- (4,3) -- cycle;
\draw[black!30] (0,0) -- (8,0) -- (12,3) -- (4,3) -- cycle;
\draw[black!60] (2,1.5) -- (10,1.5);
\draw[thick,green] (4,0) -- (6,1.5);
\draw[thick,red] (6,1.5) -- (8,3);
\draw[thick,blue] (7.5,1.5) -- (10,1.5);
\draw[thick,blue,dashed] (2,1.5) -- (4.5,1.5);
\filldraw[black] (8.63,1.6) circle (2pt);
\filldraw[black] (8.37,1.4) circle (2pt);
\node at (8.63,1.9) {$\xi_3$};
\node at  (8.37,1.1) {$\xi_2$};
\filldraw[black] (7.2,2.4) circle (2pt);
\filldraw[black] (4.8,0.6) circle (2pt);
\node at (6.8,2.5) {$\bar\xi_4$};
\node at  (5.2,0.6) {$\bar\xi_3$};
\filldraw[purple] (6.93,1.6) circle (2pt);
\filldraw[purple] (6.67,1.4) circle (2pt);
\node at (6.93,1.9) {$\xi_3$};
\node at  (6.67,1.1) {$\xi_4$};
\end{scope}
\end{tikzpicture}
\caption{\footnotesize The replica geometry $\Sigma_2$, the double cover of the $\xi$ plane joined by square root branch cuts along $A$ and its mirror in the imaginary axis. The positions of the chiral points are shown for $w\in[0,a]$ in purple and $w\in A$ in black.}
\label{fig10} 
\end{center}
\end{figure}
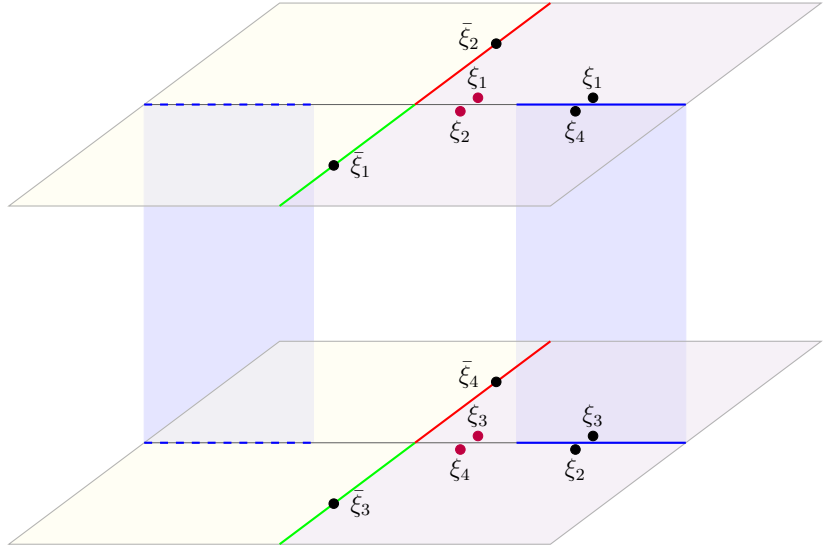

The entropy is given in terms of a four-point function on the replica geometry $\Sigma_2$ shown in figure \ref{fig10}. This can be conformally mapped via the uniformization map \eqref{nip} to the complex $z$ plane
\EQ{
S_2^{\ket{{\mathscr P}{\cal O}}}(A,w\notin A\cup P,\bar w)=-\log\Big[\prod_{j=1}^4\left|\frac{dz_j}{d\xi_j}\right|^{2\alpha^2}\big\langle\prod_{j=1}^4{\cal O}(z_j,\bar z_j)\big\rangle_{\mathbb C}\Big]\ .
}
We emphasise that in Lorentzian signature that a quantity like $|d\xi/dz|^2=d\xi/dz\cdot d\bar\xi/d\bar z$ and $d\xi/dz$ is not the complex conjugate of $d\bar\xi/dz$. Once again, even though in the BCFT doubled formalism chiral and anti-chiral have non-trivial OPEs these never are needed because the chiral operators are in close pairs and their OPE dominates. In the present case $w\in[0,a]$, these are the pairs $(12)(34)$, shown as the purple pairs in figure \ref{fig10}, which, to use the notation of section \ref{s2.3}, is the unity element of the symmetric group $\Be$. Hence we may use the formula \eqref{gug} for the correlator. In the sum over the $e_j$, there are four terms which dominate because they have inverse powers of $z_{12}$ and $z_{34}$. These are the terms with $e_1=-e_2$ and $e_3=-e_4$. Actually pairs of these terms are equal on account of the symmetry $e_j\to-e_j$ and so there are only two distinct terms:
\EQ{
\big\langle\prod_{j=1}^4{\cal O}(z_j,\bar z_j)\big\rangle_{\mathbb C}\longrightarrow \frac{2{\cal N}^4}{|z_{12}z_{34}|^{2\alpha^2}}\Big(\Big|\frac{z_{13}z_{24}}{z_{14}z_{23}}\Big|^{2\alpha^2}+\Big|\frac{z_{13}z_{24}}{z_{14}z_{23}}\Big|^{-2\alpha^2}\Big)\ ,
}
where the cross ratios
\EQ{
\frac{z_{13}z_{24}}{z_{14}z_{23}}=1\ ,\qquad  \frac{\bar z_{13}\bar z_{24}}{\bar z_{14}\bar z_{23}}=\frac1{\cos^2\gamma}\ ,
\label{piw}
}
the former following from $z_1\to z_2$ and $z_3\to z_4$. Now we assemble all the pieces to compute the entropy. It is useful to notice that the contributions from the chiral parts cancel out against the chiral contribution to the normalization \eqref{er1} because we are working in the limit $\xi_1\to\xi_2$ and $\xi_3\to\xi_4$; hence,
\EQ{
\xi_{12}\frac{dz_1}{d\xi_1}=z_{12}\ ,\qquad \xi_{34}\frac{dz_3}{d\xi_3}=z_{34}\ .
}
where ${dz_1}/{d\xi_1}={dz_2}/{d\xi_2}$ and  ${dz_3}/{d\xi_3}={dz_4}/{d\xi_4}$. The Jacobian factors for the anti-chiral points are 
\EQ{
\frac{d\bar z_j}{d\bar \xi_j}=\pm\frac{\sqrt ae^{\mp i\gamma}}{\cos^2\gamma}\ ,
\label{er2}
}
for $j=1,3$ and $2,4$, respectively. Finally we have,
\EQ{
S_2^{\ket{{\mathscr P}{\cal O}}}(A,w\notin A\cup P,\bar w=-a\tan^2\gamma)_\text{reg}=\log\frac2{1+\cos^{4\alpha^2}\gamma}\ .
\label{peb}
}

Now we repeat the steps for the case (II) with $w\in A$. In this case, under the uniformization map \eqref{nip}, the chiral points are at
\EQ{
z_1=\lambda+\delta z_1\ ,\quad z_2=-\lambda+\delta z_2\ ,\quad z_3=-\lambda+\delta z_3\ ,\quad z_4=\lambda+\delta z_4\ .
}
with $\delta z_j$ as in \eqref{pun}. In this case the chiral points are in the channel $(14)(32)$ corresponding to the element of the symmetric group $\Beta$, shown as the black pairs in the figure \ref{fig10}. In the expression for the correlator \eqref{gug} the dominant terms in the sum now have $e_1=-e_4$ and $e_3=-e_2$, 
\EQ{
\big\langle\prod_{j=1}^4{\cal O}(z_j,\bar z_j)\big\rangle_{\mathbb C}\longrightarrow \frac{2{\cal N}^4}{|z_{14}z_{32}|^{2\alpha^2}}\Big(\Big|\frac{z_{13}z_{24}}{z_{12}z_{34}}\Big|^{2\alpha^2}+\Big|\frac{z_{13}z_{24}}{z_{12}z_{34}}\Big|^{-2\alpha^2}\Big)\ ,
}
where the cross ratios
\EQ{
\frac{z_{13}z_{24}}{z_{12}z_{34}}=1\ ,\qquad  \frac{\bar z_{13}\bar z_{24}}{\bar z_{12}\bar z_{34}}=\frac1{\sin^2\gamma}\ .
\label{piw}
}
the former following from $z_1\to z_4$ and $z_3\to z_2$. Assembling all the pieces to compute the entropy, the chiral parts cancel out again:
\EQ{
\xi_{14}\frac{dz_1}{d\xi_1}=z_{14}\ ,\qquad \xi_{32}\frac{dz_3}{d\xi_3}=z_{32}\ .
}
Finally we have,
\EQ{
S_2^{\ket{{\mathscr P}{\cal O}}}(A,w\in A,\bar w=-a\tan^2\gamma)_\text{reg}=\log\frac2{1+\sin^{4\alpha^2}\gamma}\ .
\label{pea}
}

We can now use the results \eqref{peb} and \eqref{pea} to compute the R\'enyi entropy, in particular the  $n=2$ version of the UV safe entropy difference \eqref{fat}
\EQ{
\Delta S_2^{\ket{{\mathscr P}{\cal O}}}(\bar w)=\log\frac{(a-\bar w)^{2\alpha^2}+a^{2\alpha^2}}{(a-\bar w)^{2\alpha^2}+(-\bar w)^{2\alpha^2}}\ 
\label{fat2}
}
where we used $\bar w=-a\tan^2\gamma$. At the start of the trajectory, $\bar w=0$, the entropy difference equals $\log 2$ and as $\bar w$ goes into $P$ sufficiently far $\bar w\ll-a$, it goes to $-\log2$. The fact that the efficacy of teleportation increases as the ratio $|\bar w|/a$ increases is to be expected because this corresponds to the gap between $P$ and $A$ becoming smaller. In summary, we find the expected behaviour for teleportation but, of course, this is only for the second R\'enyi entropy.

\subsection{Arbitrary $n$}\label{s3.2}

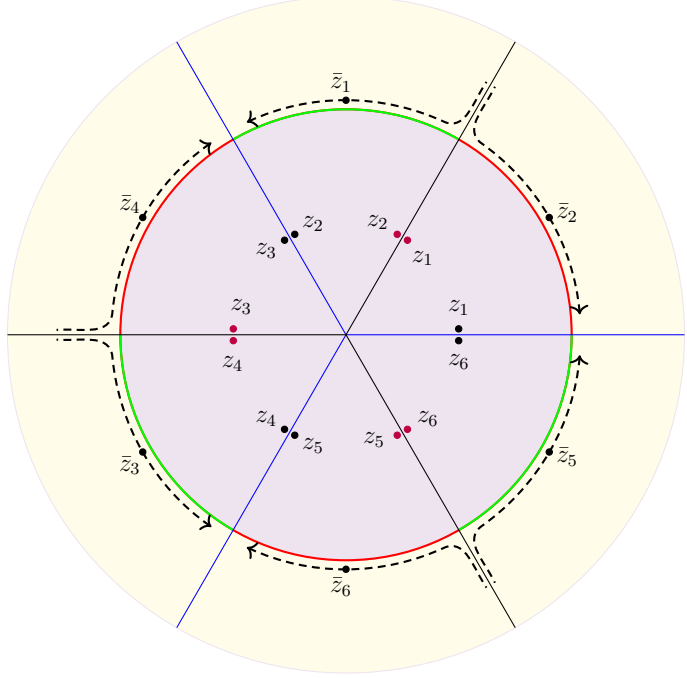
\begin{figure}[ht]
\begin{center}
\begin{tikzpicture} [scale=0.6,every node/.style={scale=0.8}]
\draw[fill = yellow!10!white, draw = Plum!10!white] (0,0) circle (7.5cm);
%\draw[fill = white, draw = white] (0,0) circle (5cm);
\draw[fill = Plum!10!white, draw = Plum!10!white] (0,0) circle (5cm);
\draw[red,thick] (0,0) circle (5cm); 
\draw[green,thick] (60:5) arc[radius=5, start angle=60, end angle= 120];
\draw[green,thick] (180:5) arc[radius=5, start angle=180, end angle= 240];
\draw[green,thick] (0:5) arc[radius=5, start angle=0, end angle= -60];
\draw[blue] (0,0) -- (0:7.5cm);
\draw[blue] (0,0) -- (120:7.5cm);
\draw[blue] (0,0) -- (240:7.5cm);
\draw (0,0) -- (60:7.5cm);
\draw (0,0) -- (180:7.5cm);
\draw (0,0) -- (-60:7.5cm);
\draw[<-,thick,densely dashed] (0,0) ++(5:5.2cm) arc (5:55:5.2cm) to[out=140,in=-115] (3.3,5.5);
\draw[<-,thick,densely dashed] (0,0) ++(-5:5.2cm) arc (-5:-55:5.2cm) to[out=-140,in=115] (3.3,-5.5);
\filldraw[black] (0,0) ++(30:5.2cm) circle (2pt);
\filldraw[black] (0,0) ++(-30:5.2cm) circle (2pt);
\filldraw[black] (0,0) ++ (3:2.5cm) circle (2pt);
\filldraw[black] (0,0) ++ (-3:2.5) circle (2pt);
\filldraw[purple] (0,0) ++ (57:2.5cm) circle (2pt);
\filldraw[purple] (0,0) ++ (63:2.5) circle (2pt);
%
%\filldraw[black] (3.3,5.5) circle (2pt);
%\filldraw[black] (3.3,-5.5) circle (2pt);
%\filldraw[black] (5.2,0.2) circle (2pt);
%\filldraw[black] (5.2,-0.2) circle (2pt);
%
\node at (2.5,0.6) {$z_1$};
\node at (2.5,-0.6) {$z_6$};
\node at (1.7,1.7) {$z_1$};
\node at (0.7,2.5) {$z_2$};
\node at (4.9,2.7) {$\bar z_2$};
\node at (4.9,-2.7) {$\bar z_5$};
\begin{scope}[rotate=120]
\draw[<-,thick,densely dashed] (0,0) ++(5:5.2cm) arc (5:55:5.2cm) to[out=140,in=-115] (3.3,5.5);
\draw[<-,thick,densely dashed] (0,0) ++(-5:5.2cm) arc (-5:-55:5.2cm) to[out=-140,in=115] (3.3,-5.5);
\filldraw[black] (0,0) ++(30:5.2cm) circle (2pt);
\filldraw[black] (0,0) ++(-30:5.2cm) circle (2pt);
\filldraw[black] (0,0) ++ (3:2.5cm) circle (2pt);
\filldraw[black] (0,0) ++ (-3:2.5) circle (2pt);
\filldraw[purple] (0,0) ++ (57:2.5cm) circle (2pt);
\filldraw[purple] (0,0) ++ (63:2.5) circle (2pt);
%
%
%\filldraw[black] (3.3,5.5) circle (2pt);
%\filldraw[black] (3.3,-5.5) circle (2pt);
%\filldraw[black] (5.2,0.2) circle (2pt);
%\filldraw[black] (5.2,-0.2) circle (2pt);
%
\node at (2.5,0.6) {$z_3$};
\node at (2.5,-0.6) {$z_2$};
\node at (1.7,1.7) {$z_3$};
\node at (0.7,2.5) {$z_4$};
\node at (4.9,2.7) {$\bar z_4$};
\node at (4.9,-2.7) {$\bar z_1$};
\end{scope}
\begin{scope}[rotate=-120]
\draw[<-,thick,densely dashed] (0,0) ++(5:5.2cm) arc (5:55:5.2cm) to[out=140,in=-115] (3.3,5.5);
\draw[<-,thick,densely dashed] (0,0) ++(-5:5.2cm) arc (-5:-55:5.2cm) to[out=-140,in=115] (3.3,-5.5);
\filldraw[black] (0,0) ++(30:5.2cm) circle (2pt);
\filldraw[black] (0,0) ++(-30:5.2cm) circle (2pt);
\filldraw[black] (0,0) ++ (3:2.5cm) circle (2pt);
\filldraw[black] (0,0) ++ (-3:2.5) circle (2pt);
\filldraw[purple] (0,0) ++ (57:2.5cm) circle (2pt);
\filldraw[purple] (0,0) ++ (63:2.5) circle (2pt);
%
%
%\filldraw[black] (3.3,5.5) circle (2pt);
%\filldraw[black] (3.3,-5.5) circle (2pt);
%\filldraw[black] (5.2,0.2) circle (2pt);
%\filldraw[black] (5.2,-0.2) circle (2pt);
%
\node at (2.5,0.6) {$z_5$};
\node at (2.5,-0.6) {$z_4$};
\node at (1.7,1.7) {$z_5$};
\node at (0.7,2.5) {$z_6$};
\node at (4.9,2.7) {$\bar z_6$};
\node at (4.9,-2.7) {$\bar z_3$};
\end{scope}
\end{tikzpicture}
\caption{\footnotesize The uniformatization to the $z$ cylinder for the case $n=3$. The 3 replicas correspond to the 3 wedges bounded by the blues lines, the cuts that have been opened up. The boundary of the slit is mapped to the unit circle. The black lines correspond to the image of the gap between $P$ and $A$, $w\in[0,a]$. The yellow region corresponds the image region $\RE\xi<0$ of the doubled formalism. The trajectories of the anti-chiral coordinates $\{\bar z_i\}$ and replica images as they probe deep into $P$ are shown. }
\label{fig5} 
\end{center}
\end{figure}

We now generalize the result of the last section to arbitrary $n$. A suitable uniformization map is
\EQ{
z^n=\frac{\xi-\sqrt a}{\xi+\sqrt a}\ .
}
As in the $n=2$ case, the boundary in the $\xi$ plane $\RE\xi=0$ is the unit circle $|z|=1$. The uniformized $z$ plane is shown in figure \ref{fig5} for the case $n=3$. The images of the anti-chiral points are at
\EQ{
\bar z_{2j-1}=\exp\Big[\frac in(2\gamma+\pi(2j-1))\Big]\ ,\quad \bar z_{2j}=\exp\Big[\frac in(-2\gamma+\pi(2j-1))\Big]\ .
\label{pic}
}

We will concentrate on computing the entropy difference \eqref{fat} directly. This greatly simplifies the computation because the normalization and conformal map factors cancel out to leave a simple ratio of correlators on the $z$ plane
\EQ{
\Delta S_n^{\ket{{\mathscr P}{\cal O}}}(\bar w=-a\tan^2\gamma)=\frac1{1-n}\log\frac{\big\langle \prod_{j=1}^{2n}{\cal O}(z_j,\bar z_j)\big\rangle_\Beta}{\big\langle \prod_{j=1}^{2n}{\cal O}(z_j,\bar z_j)\big\rangle_\Be}\ .
}
The labels $\Beta$ and $\Be$ indicate the chiral points are in the corresponding channels, so $z_{2j+1}\to z_{2j}$ and $z_{2j-1}\to z_{2j}$, respectively. Using the formula \eqref{gug} for the correlators, up to terms that cancel in the ratio, for the $\Beta$ channel, the terms with $\hat e_j\equiv e_{2j+1}=-e_{2j}$ dominate and we have
\EQ{
\big\langle\cdots\rangle_\Beta\thicksim\prod_{j=1}^n\bar z_{2j+1,2j}^{-1}\sum_{\hat e_j=\pm1}\prod_{j<k=1}^n\Big(\frac{\bar z_{2j+1,2k+1}\bar z_{2j,2k}}{\bar z_{2j+1,2k}\bar z_{2j,2k+1}}\Big)^{\alpha^2\hat e_j\hat e_k}\ ,
\label{gux}
}
On the other hand, for the $\Be$ channel it is the terms with $\hat e_j\equiv e_{2j-1}=-e_{2j}$ that dominate and
\EQ{
\big\langle\cdots\rangle_\Be\thicksim\prod_{j=1}^n\bar z_{2j-1,2j}^{-1}\sum_{\hat e_j=\pm1}\prod_{j<k=1}^n\Big(\frac{\bar z_{2j-1,2k-1}\bar z_{2j,2k}}{\bar z_{2j-1,2k}\bar z_{2j,2k-1}}\Big)^{\alpha^2\hat e_j\hat e_k}\ .
\label{guk}
}
(labels defined mod $n$).

The ratio of \eqref{guk} and \eqref{gux} can be evaluated for some low values of $n$ and the specific choice $\alpha=1$. For $n=3$, we find
\EQ{
\Delta S_3^{\ket{{\mathscr P}{\cal O}}}(\bar w=-a\tan^2\gamma)=\frac12\log\frac AB\ ,
}
where
{\small\EQ{
A&=2\Big(23+16\cos\frac{4\gamma}3+10\cos\frac{8\gamma}3+4\cos4\gamma+\cos\frac{16\gamma}3\Big)\ ,\\[5pt]
B&=46-16\cos\frac{4\gamma}3-10\cos\frac{8\gamma}3+8\cos4\gamma-\cos\frac{16\gamma}3\\[5pt] &\qquad\qquad\qquad-16\sqrt3\sin\frac{4\gamma}3+10\sqrt3\sin\frac{8\gamma}3-\sqrt3\sin\frac{16\gamma}3\ .
}}

Then for $n=4$, we find
\EQ{
\Delta S_4^{\ket{{\mathscr P}{\cal O}}}(\bar w=-a\tan^2\gamma)=\frac13\log\frac AB\ ,
}
where
{\small\EQ{
A&=2331+1880\cos\gamma+1920\cos2\gamma+968\cos3\gamma+732\cos4\gamma+216\cos5\gamma\\[5pt] &\qquad\qquad\qquad+128\cos6\gamma+8\cos7\gamma+9\cos8\gamma\,\\[5pt]
B&=2331-1920\cos2\gamma+732\cos4\gamma-128\cos6\gamma+9\cos8\gamma-1880\sin\gamma\\[5pt] &\qquad\qquad\qquad+968\sin3\gamma-216\sin5\gamma+8\sin7\gamma\ .
}}

\pgfdeclareimage[interpolate=true,width=9cm]{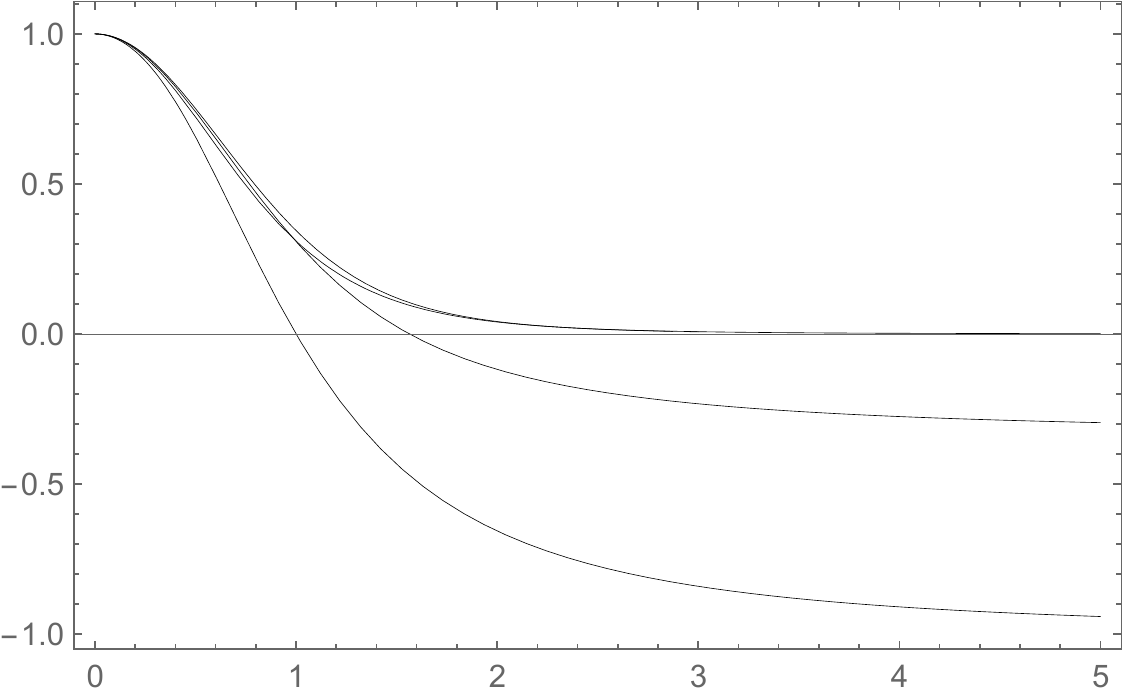}{pic3}
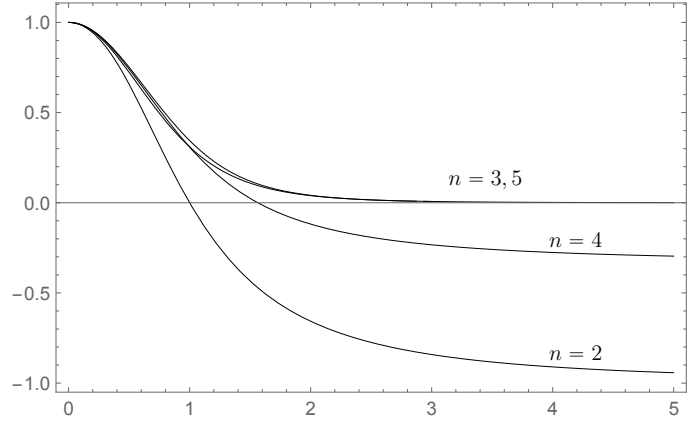
\begin{figure}
\begin{center}
\begin{tikzpicture}[scale=1,every node/.style={scale=0.7}]
\pgftext[at=\pgfpoint{0cm}{0cm},left,base]{\pgfuseimage{pic3}} 
\node at (7.5,0.9) {$n=2$};
\node at (7.5,2.4) {$n=4$};
%\node (a1) at (5,3.3) {$n=3$};
\node (a2) at (6.3,3.2) {$n=3,5$};
\end{tikzpicture}
\caption{\footnotesize $\Delta S_n^{\ket{{\mathscr P}{\cal O}}}(A)$ as a function of $\tan\gamma =\sqrt{|\bar w|/a}$ for $n=2,3,4,5$ in units of $\log2$. For $\bar w=0$ ($\gamma=0$) when the anti-chiral  point is just outside $\PR$, $\Delta S^{\ket{{\mathscr P}{\cal O}}}_n(A)=\log2$ and when the anti-chiral  point is deep inside $\PR$, $\bar w\ll-a$ ($\gamma\to\frac\pi2$), the result depends on $n$ as in \eqref{rws}.}
\label{fig6}
\end{center}
\end{figure}

In figure \ref{fig6} we plot $\Delta S_n^{\ket{{\mathscr P}\cal O}}(A)$ as a function of $|\bar w|/a=\tan^2\gamma$ for $n=2,3,4,5$ in units of $\log 2$. In all cases, as $\gamma\to0$, $\Delta S_n^{\ket{{\mathscr P}{\cal O}}}(A)\to \log2$, precisely the entropy of the operator quench. We now explain this in a universal way.

\subsection{Universal limits}

In the limit $\gamma\to0$, the anti-chiral points are in the $\bar z_{2j-1}\to\bar z_{2j}$, i.e.~$\Be$, channel. Hence, using the general result \eqref{get} we deduce that the R\'enyi excess entropy
\EQ{
S_n^{\ket{{\mathscr P}{\cal O}}}(A,w\in A,\bar w=0)_\text{reg}=\frac{{\mathscr D}(\Beta,\Be)}{n-1}\log d_{\cal O}=\log d_{\cal O}\ ,
} 
and 
\EQ{
S_n^{\ket{{\mathscr P}{\cal O}}}(A,w\notin A,\bar w=0)_\text{reg}=\frac{{\mathscr D}(\Be,\Be)}{n-1}\log d_{\cal O}=0\ . 
}
We can also understand the limit where the anti-chiral points are well into the projection region, $\bar w\ll-a$, i.e.~$\gamma\to\frac\pi2$. In this limit, it is clear from \eqref{pic} that the anti-chiral points are paired in the channel $(2j-3,2j)$ which is illustrated in figure \ref{fig5} for $n=3$. This channel corresponds is also defined by a cyclic permutation by now the inverse $\Bsigma=\Beta^{-1}$. Hence, in this limit,
\EQ{
S_n^{\ket{{\mathscr P}{\cal O}}}(A,w\in A,\bar w\ll -a)_\text{reg}&\to \frac{{\mathscr D}(\Beta,\Beta^{-1})}{n-1}\log d_{\cal O}\\[5pt] &=\log d_{\cal O}\times \begin{cases}\frac{n-2}{n-1} & n\text{ even}\ ,\\ 1 & n\text{ odd}\ .\end{cases}\ ,
\label{pwina}
} 
and 
\EQ{
S_n^{\ket{{\mathscr P}{\cal O}}}(A,w\not\in A\cup P,\bar w\ll -a)_\text{reg}\to \frac{{\mathscr D}(\Be,\Beta^{-1})}{n-1}\log d_{\cal O}=\log d_{\cal O}\ . 
}
In the above, notice that ${\mathscr D}(\Beta,\Beta^{-1})=n-1$ for $n$ odd because in that case $\Beta^2$ has one cycle; however, for $n$ even it equals $n-2$ because $\Beta^2$ has 2 cycles.  It is surprising that \eqref{pwina} is non-vanishing for all $n$ except $n=2$. For efficient projection we would expect the R\'enyi excess to vanish when the state is teleported to $A$.

To summarize, we have the universal limits, 
\EQ{
\Delta S_n^{\ket{{\mathscr P}{\cal O}}}(\bar w\to 0)&=\log d_{\cal O}\ ,\\[5pt]
\Delta S_n^{\ket{{\mathscr P}{\cal O}}}(\bar w\ll -a)&=\begin{cases}\frac{\log d_{\cal O}}{1-n} & n\text{ even}\ ,\\ 0 & n\text{ odd}\ .\end{cases}
\label{rws}
}
It is clear that teleportation is not maximally efficient, as we would have expected since the projection is not especially tuned to the vacuum state of the CFT. It is also apparent that there is rather unusual $n$ dependence.

\subsection{Analytically continuing $n\to1$}

Having computed the R\'enyi entropies, we now address whether we can take the analytic continuation $n\to1$ to extract the mutual information $I(V,A)$. Na\"\i vely, one could expect this to be the $n\to1$ limit of 
\EQ{
I_n=S(V)-\Delta S_n^{\ket{{\mathscr P}{\cal O}}}(\bar w)\ . 
}
One systematic way to take the von Neumann limit of R\'enyi entropies is described in \cite{DHoker:2020bcv}. When the R\'enyi entropies $S_n$ are known for integer values $n=2,3,\ldots$ one defines a generating function
\EQ{
G(z)=\sum_{n=1}^\infty\frac{z^n}n\big( e^{-nS_{n+1}}-1\big)\ .
} 
In a finite dimensional setting $G(z)$ is analytic everywhere apart from branch points on the real axis $z\geq1$. In that case, the von Neumann entropy is obtained in the limit along the negative real axis
\EQ{
S=\lim_{z\to-\infty}G(z)\ .
}
The procedure works in the context of one and two intervals in a CFT with a UV regulator and has been conjectured to apply also to mutual informations  \cite{DHoker:2020bcv}. If we attempt to apply it to the mutual information in the limit $\bar w\ll -a$ that follow from \eqref{rws}, that is
\EQ{
I_n=\begin{cases}\frac n{n-1}\log d_{\cal O} & n\text{ even}\ ,\\ \log d_{\cal O} & n\text{ odd}\ ,
\end{cases}
}
then we find that the generating function equals
\EQ{
G(z)=\log d_{\cal O}+\frac1{2d_{\cal O}}\log\frac{d_{\cal O}+z}{d_{\cal O}-z}-\frac12\log(d_{\cal O}^2-z^2)+\log(1-z)\ .
}
A problem now emerges, $G(z)$ has a branch cut that ends on the real  axis at $z=-d_{\cal O}$ (and another that ends at $z=d_{\cal O}$). Analytically continuing past the branch point yields an ambiguous imaginary part. If we simply ignore this imaginary part, we have for $\bar w\ll- a$,
\EQ{
I(V,A)=\log d_{\cal O}\ .
}
We leave a full understanding of the $n$ dependence to future work.

\section{Conformal blocks and R\'enyi entropies}\label{s4}

An alternative, but completely equivalent, approach to computing R\'enyi entropies in CFTs is to replicate the CFT itself rather than the background geometry. So the background geometry in our case is the $w$ complex plane. The end points of $A$ at $w=a$ and $\infty$, where the replicas are branched, are then represented in the replicated CFT by twist operators \cite{Asplund:2014coa, Kusuki:2017upd}. The replicated CFT has central charge $nc$, and the replicated quench operator 
\begin{equation}
{\cal O}_{ n} \equiv \underbracket{{\cal O}\otimes {\cal O}\otimes{\cal O}\otimes\ldots {\cal O}}_{n\,{\rm times}}\ ,
\end{equation}
has conformal dimension $2n\Delta$. The R\'enyi entropies of the system $A$, given by our semi-infinite interval $w\in [a,\infty]$ in the CFT,  are then computed by the correlator
\EQ{
S_{n}^{\ket{\cal O}}(A)=\frac{1}{1-n}\log \,{\cal C}_n\ ,
}
where
\EQ{
{\cal C}_n=\frac{\langle{\cal O}^{\dagger}_{n}(w_2, \bar w_2) \,\sigma_n(a,\bar a)  \,{\cal O}_{n}(w_1, \bar w_1) \rangle_{{\mathscr P}}}{\langle{\cal O}^{\dagger}_{n}(w_2, \bar w_2) {\cal O}_{n}(w_1, \bar w_1) \rangle_{\mathscr P} \,\,\langle\sigma_n(a,\bar a) \rangle_{{\mathscr P}}}\ ,
\label{threept}
}
in the projected state.\footnote{ In the BCFT picture on the $\xi$-plane, the R\'enyi entropy of the semi-infinite interval $[\sqrt{a},\infty]$ is the same as that of its complement $[\sqrt a, 0]$ including the boundary.} Here $\sigma_n$ is a twist field viewed as a primary CFT field with conformal dimension $2h_n$,
\EQ
{
h_n = \frac{c}{24}\left(n-\frac1n\right)\,.}
While three-point functions of primary fields are fixed by conformal invariance for  a CFT in the vacuum state, the presence of the projector makes \eqref{threept} nontrivial  and non-universal.  The large-$c$ limit however provides a route to compute the desired correlation function in the projected state. For fixed integer $n\neq 1$,  twist fields can be viewed as heavy operators whose conformal dimensions scale with the central charge,  whilst the operators creating the local quench are light with parametrically small scaling dimensions,
\EQ
{
\frac{\Delta}{c} \ll 1\,, \qquad \frac{h_n}{c}= O(c^0)\,,
}
in the $c\to\infty$ limit.

In previous sections we have exploited the fact that the projected state $\ket{B}$ is a Cardy state and the correlation functions in this state can be viewed as BCFT correlators on the half-plane. In this section, we will adopt a slight generalisation of the map \eqref{openup} which opens up the projection slit, so that the CFT  can also be considered at a finite temperature $\beta^{-1}$:
\EQ{
\xi = \sqrt{\frac{\beta}{2\pi}}\left(1- e^{-2\pi w/\beta}\right)^{\frac12}\ ,
\qquad -\frac{\beta}{2}\leq {\rm Im}\, w \leq \frac{\beta}{2}\ ,
\label{mapbeta} 
}
with shifts of $w$ along the imaginary axis identified according to $w\sim w +i\beta$. 
In the zero temperature limit $\beta\to \infty$, this reduces to the map  $\xi =\sqrt w$.

For theories with finite central charge, as illustrated in the example considered earlier, the jumps in  R\'enyi entropies are independent of the original state of the theory for vanishingly small pulse widths $\epsilon$ when the relevant correlators are dominated by competing OPE channels. In the large-$c$ theory however, as we explain below, results will depend non-trivially on pulse widths and so it is instructive to examine different initial states on which the projection acts. The new map \eqref{mapbeta}  takes the thermal cylinder  with the projection slit to the half-plane ${\rm Re}\, \xi >0$.  The normalised correlator \eqref{threept} on the half-plane is now,
\EQ
{
{\cal C}_n=\frac{\langle{\cal O}^{\dagger}_n(\xi_2, \bar \xi _2) \,\sigma_n({\mathfrak a},\bar {\mathfrak a}) \,{\cal O}_{n}(\xi_1, \bar \xi_1) \rangle_{\rm BCFT}}{\langle{\cal O}^{\dagger}_n(\xi_2, \bar \xi_2) {\cal O}_{n}(\xi_1, \bar \xi_1) \rangle_{\rm BCFT} \,\,\langle\sigma_n({\mathfrak a},\bar {\mathfrak a}) \rangle_{\rm BCFT}}\,,\label{fourptbcft}
} 
where $(\xi_i, \bar \xi_i)$ are images on the half-plane of the points $(w_i, \bar w_i)$ on the cylinder, and 
\EQ
{ {\mathfrak a} = \bar {\mathfrak a} = \sqrt{\frac{\beta}{2\pi}}\left(1-e^{-2\pi a /\beta}\right)^{\frac12}\,.\label{mapa}
}
In the absence of the local quench, the twist field $\sigma_n$ acquires a one-point function in the projected state, which determines the R\'enyi entropy of $A$ in this state,\footnote{The one-point function in the projected vacuum arises through the one-point function of the twist operator in BCFT,
\EQ{
\langle\sigma_n( a, \bar{a})\rangle_{\mathscr P}= \left|\frac{\partial\xi}{\partial w}\right|^{2h_n}_{w=\bar w = a}\langle\sigma_n( {\mathfrak a}, \bar{\mathfrak a})\rangle_{\rm BCFT}\,,\qquad \langle\sigma_n({\mathfrak a}, \bar{\mathfrak a})\rangle_{\rm BCFT} = g_b^{1-n}\left(\frac{\delta}{2{\mathfrak a}}\right)^{2 h_n}\,.
} }
\begin{eqnarray}
&&\langle\sigma_n( a, \bar{a})\rangle_{\mathscr P}=g_b^{1-n}\left(\frac{\pi\delta}{2\beta}\right)^{2h_n}\left(e^{2\pi a/\beta}-1\right)^{-2 h_n}\\\nonumber\\
&&S_n^{\ket{{\mathscr P}{\cal O}}}(A)=\frac{c}{12}\cdot\frac{n+1}{n}\log\left[\frac{2\beta}{\delta \pi}\left(e^{2\pi a/\beta}-1\right)\right]\,,
\end{eqnarray}
where $\delta$ is a UV cutoff and $g_b$ the boundary entropy for the BCFT.
On the half-plane, we can employ the doubling trick to view the  three-point correlation function  as a  six-point chiral correlator on the plane with {\em two} heavy (twist) fields. This is not universal, but we can analyse it in a large-$c$ semiclassical limit assuming identity conformal block  dominance. 

\subsection{Six-point chiral correlator}

\paragraph{Integer $n$ and HHLLLL limit:} For integer-valued $n$, the twist fields are heavy operators in the large-$c$ limit, and the  chiral six-point correlator on the $\xi$-plane thus  involves  four light insertions  and two heavy ones, 
\EQ
{
{\cal C}_n =\sum_{a, b}\frac{\langle V_{n, a}^{\dagger}(\xi_2) \,\bar V_{n, a}^\dagger(- \bar\xi _2) \,\sigma_n({\mathfrak a})\,\bar \sigma_n(-{\mathfrak a}) \,V_{n, b}(\xi_1) \,\bar V_{n, b}(- \bar\xi _1) \rangle}{\langle{\cal O}^{\dagger}_n(\xi_2, \bar \xi_2) \,{\cal O}_{n}(\xi_1, \bar \xi_1) \rangle \,\,\langle\sigma_n({\mathfrak a}) \,\bar \sigma_n(- {\mathfrak a}) \rangle}\,.\label{sixpoint}
}
The chiral decomposition of the quench operator ${\cal O}_n$ is formally shown above, but it is worth recalling that the quantum dimension $d_{\cal O}$ in the large-$c$ setting is formally infinite, potentially nonperturbatively in $c$ \cite{Caputa:2014vaa}.  
\begin{figure}[ht]
\begin{center}
\begin{tikzpicture}[scale=0.7,every node/.style={scale=0.8},decoration={markings,mark=at position 0.5 with {\arrow{>}}}]
\draw[very thick] (-2,-1) -- (-1,0);
\draw[very thick] (-1,0) -- (-2,1);
\draw[very thick,postaction] (5,1) -- (4,0);
\draw[very thick] (4,0) -- (5,-1);
\draw[very thick] (-1,0) -- (4,0);
\node[left] at (-1.4,-1.4) {$\bar V^\dagger_n(-\bar \xi_2)$};
\node[left] at (-1.4,1.4) {$V^\dagger_n (\xi_2)$};
\node[right] at (4.2,-1.3) {$ \bar V_n(-\bar \xi_1)$};
\node[right] at (4.2,1.3) {$V_n(\xi_1)$};
%
%\draw[<->,line width=1.5mm] (5.5,0) -- (8,0);
%
%\begin{scope}[xshift=11cm,yshift=-1.2cm]
%\draw[very thick] (1,-2) -- (1,-2);
%\draw[very thick] (1,-1) -- (2,-2);
\draw[very thick] (2.5,3) -- (1.5,2);
\draw[very thick] (1.5,2) -- (0.5,3);
\draw[very thick] (1.5,0) -- (1.5,2);
\node[left] at (0.9,3.3) {$\sigma_n({\mathfrak a})$};
\node[right] at (2.1,3.3) {$\bar \sigma_n(-{\mathfrak a})$};
%\draw[very thick] (7,2) -- (6,1);
%\draw[very thick] (6,1) -- (5,2);
%\draw[very thick] (6,0) -- (6,1);
%\node[left] at (5.4,2.3) {$\bar \sigma_n(1)$};
%\node[right] at (6.6,2.3) {$\sigma_n(-1)$};
%\end{scope}
\end{tikzpicture}
\caption{\footnotesize  OPE channel in which twist fields and their BCFT images  on the plane fuse in pairs in the presence of the chiral quench operator insertions in the $\xi$-plane,  ${\cal O}_n(\xi,\bar \xi)\to \sum_a V_{n, a}(\xi) \bar V_{n, a}(-\bar \xi)$.}
\label{hhllll}
\end{center}
\end{figure}
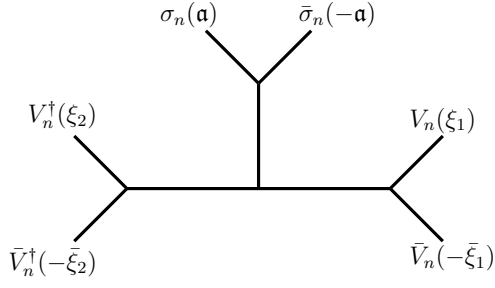
The vacuum block contribution to such  correlators has been computed in \cite{Anous:2019yku}. In our application, it is dominated by the OPE channel in which each chiral twist field fuses with its BCFT image  as shown in  figure \ref{hhllll}. The corresponding result  for the leading terms in a $\Delta/c$-expansion is,
\EQ{
{\cal C}_n(u,v) = \left(\alpha^2\frac{u^{1/2} - u^{-1/2}}{u^{\alpha/2}-u^{-\alpha/2}}\cdot \frac{v^{1/2} - v^{-1/2}}{v^{\alpha/2}-v^{-\alpha/2}}\right)^{2n \Delta} \left(1+ n^2\frac{\Delta^2}{c} \,\delta{\cal C}_n\right)\,,\label{hhllll}
}
where\footnote{The correction term at order $\Delta^2/c$ is,
\EQ{\delta{\cal C}_n=2(1-\chi)^2 \,{}_2F_1(2,2,4,1-\chi)\,,\qquad \chi= \frac{(1-w^\alpha)(u^\alpha-(w/v)^\alpha)}{(1-(w/u)^\alpha)(u^\alpha-w^\alpha)}\,,}
with $w= \frac{(\xi_1-{\mathfrak a})(-\bar\xi_2+{\mathfrak a})}{(\xi_1+{\mathfrak a})(-\bar\xi_2 -a)}$. This contribution  is parametrically subleading, but also remains numerically  small in all scenarios we discuss below. 
}
\EQ
{
\alpha = \sqrt{1- \frac{24}{nc}\, h_n } =\frac{1}{n}\,,
}
and the conformal cross-ratios $u$ and $v$ are defined as
\EQ
{
u=e^{i\theta_u}=\frac{(\xi_1-{\mathfrak a})(\xi_2+{\mathfrak a})}{(\xi_1+{\mathfrak a})(\xi_2-{\mathfrak a})}\,,\qquad v= e^{i\theta_v} =\frac{(-\bar\xi_1-{\mathfrak a})(-\bar\xi_2+{\mathfrak a})}{(-\bar\xi_1+{\mathfrak a})(-\bar\xi_2-{\mathfrak a})}\,.
}
Both cross-ratios have unit modulus, $|u|=|v|=1$, since the insertion points of the chiral fields satisfy $\xi_{1} = \xi_{2}^*$ and $\bar\xi_1=\bar \xi_2^*$. The R\'enyi entropy excess in the large-$c$ theory then assumes a simple form in terms of the phases $\theta_{u,v}$\,,
\EQ
{
S_n^{\ket{{\mathscr P}\cal O}}(A)= \frac{2n\Delta}{1-n}\log\left(\frac{\sin\frac{\theta_u}{2}}{n\sin \frac{\theta_u}{2n}}\cdot \frac{\sin\frac{\theta_v}{2}}{n\sin \frac{\theta_v}{2n}}\right)\,.\label{renyilargec}
}
\paragraph{von Neumann entropy:} The von Neumann limit of the R\'enyi entropies must be taken carefully since twist operators become light in the $n\to 1$ limit.  The chiral correlator \eqref{sixpoint} then has two light twist operators with $\frac{h_n}{c}= O((n-1))$ and four other chiral insertions with $\frac{\Delta}{c}=O((n-1)^0)$. Whilst \eqref{renyilargec} is well defined as $n\to 1$, it is no longer consistent to use the HHLLLL limit to evaluate \eqref{sixpoint}. Taking the quench operator dimension $\Delta$ to scale with $c$ can provide a heirarchy between the conformal dimensions but then we are left with a correlation function involving four heavy and two light operators. This does not simplify except in certain limits e.g.~when the quench operator is inserted {\em at} the boundary \cite{Kusuki:2017upd}. We will therefore restrict our discussion to R\'enyi entropies  with integer $n\neq 1$.

\subsection{Finite interval and eight-point chiral correlator} 
The analysis above, for  the semi-infinite interval $A$, also finds application in the situation with a finite interval $A=[a,b]$  in the $w$-plane. Computation of  the R\'enyi entropy of $A$ in the projected state with a local quench, requires a pair of twist field insertions (twist and anti-twist) at $\xi = {\mathfrak a}, {\mathfrak b}$, the endpoints of $A$ in the $\xi$ half-plane. This leads us to consider a four-point BCFT correlator which is then equivalent to an {\em eight}-point chiral correlator on the $\xi$-plane, involving four heavy  and four light chiral fields. This is, of course, unknown in general, but a four-point BCFT correlator is given by the sum over two OPE channels, one where the twist-antitwist pair are factorised (boundary channel) and one where they are not (bulk channel).  
\begin{figure}[ht]
\begin{center}
\begin{tikzpicture}[scale=0.7,every node/.style={scale=0.8},decoration={markings,mark=at position 0.5 with {\arrow{>}}}]
\draw[very thick] (-2,-1) -- (-1,0);
\draw[very thick] (-1,0) -- (-2,1);
\draw[very thick,postaction] (9,1) -- (8,0);
\draw[very thick] (8,0) -- (9,-1);
\draw[very thick] (-1,0) -- (8,0);
\node[left] at (-1.4,-1.4) {$\bar V^\dagger_n(-\bar \xi_2)$};
\node[left] at (-1.4,1.4) {$V^\dagger_n (\xi_2)$};
\node[right] at (8.2,-1.3) {$ \bar V_n(-\bar \xi_1)$};
\node[right] at (8.2,1.3) {$V_n(\xi_1)$};
%
%\draw[<->,line width=1.5mm] (5.5,0) -- (8,0);
%
%\begin{scope}[xshift=11cm,yshift=-1.2cm]
%\draw[very thick] (1,-2) -- (1,-2);
%\draw[very thick] (1,-1) -- (2,-2);
\draw[very thick] (2,2) -- (1,1);
\draw[very thick] (1,1) -- (0,2);
\draw[very thick] (1,0) -- (1,1);
\node[left] at (0.4,2.3) {$\sigma_n({\mathfrak a})$};
\node[right] at (1.6,2.3) {$\bar \sigma_n(-{\mathfrak a})$};
\draw[very thick] (7,2) -- (6,1);
\draw[very thick] (6,1) -- (5,2);
\draw[very thick] (6,0) -- (6,1);
\node[left] at (5.4,2.3) {$\bar \sigma_n({\mathfrak b})$};
\node[right] at (6.6,2.3) {$\sigma_n(-{\mathfrak b})$};
%\end{scope}
\end{tikzpicture}
\caption{\footnotesize  OPE channel of the eight-point chiral correlator relevant for the finite interval situation.}
\label{factorised}
\end{center}
\end{figure}
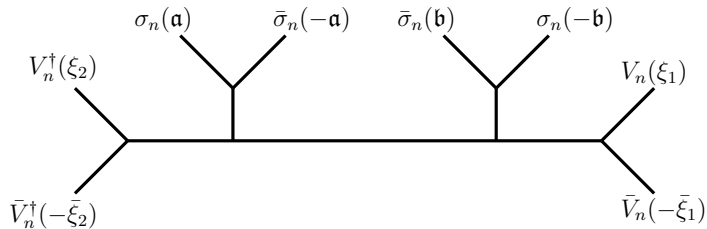
Effective teleportation of the operator quench by the projection only occurs in the {\em factorised} or boundary channel i.e. when the twist-antitwist pair are well separated ${\mathfrak a} \ll {\mathfrak b}$, leading to  non-vanishing one-point functions,
\EQ{
\langle\sigma_n({\mathfrak a}, \bar{\mathfrak a})\rangle_{\rm BCFT} = g_b^{1-n}\left(\frac{\delta}{2{\mathfrak a}}\right)^{2 h_n}\,,\qquad \langle\bar\sigma_n({\mathfrak b}, \bar{\mathfrak b})\rangle_{\rm BCFT} = g_b^{1-n}\left(\frac{\delta}{2 {\mathfrak b}}\right)^{2 h_n}\,.
}
In this limit, the eight-point chiral correlation function on the plane is dominated by the OPE channel in which each chiral twist field fuses with its BCFT image  as shown in  figure \ref{factorised}. Inserting a complete set of states and exploiting the orthogonality of two-point functions in CFT \cite{Caputa:2014eta, Kawamoto:2022etl},  the correlator is schematically a product of two six-point (HHLLLL) correlation functions,
\EQ
{
{\cal C}_n\big|_{{\mathfrak a}\ll {\mathfrak b}} \sim\langle {\cal O}^\dagger_{n} \,\sigma_n({\mathfrak a},\bar{\mathfrak a}) \, {\cal O}_{n}\rangle\,\langle {\cal O}^\dagger_{n} \,\sigma_n({\mathfrak b},\bar{\mathfrak b}) \, {\cal O}_{n}\rangle.\label{eightpt}
}
\subsection{Evolution of the R\'enyi excess}

We have noted previously that the projection does not commute with the CFT Hamiltonian, and so evaluating \eqref{renyilargec} as a function of  time or light cone coordinates $(w,\bar w)$, does not yield the actual time evolution of the local quench state  following the projection. Instead, the functional dependence  on $(w, \bar w)$ yields  the  entanglement structure of a family of post selected states with local quenches at different spatial separations. To this end, we must now understand the evolution of the phases of cross-ratios $u$ and $v$ as a function of the chiral and anti-chiral positions $(w, \bar w)$ of the quench operator.

\paragraph{Scenario I $w\in A$: }
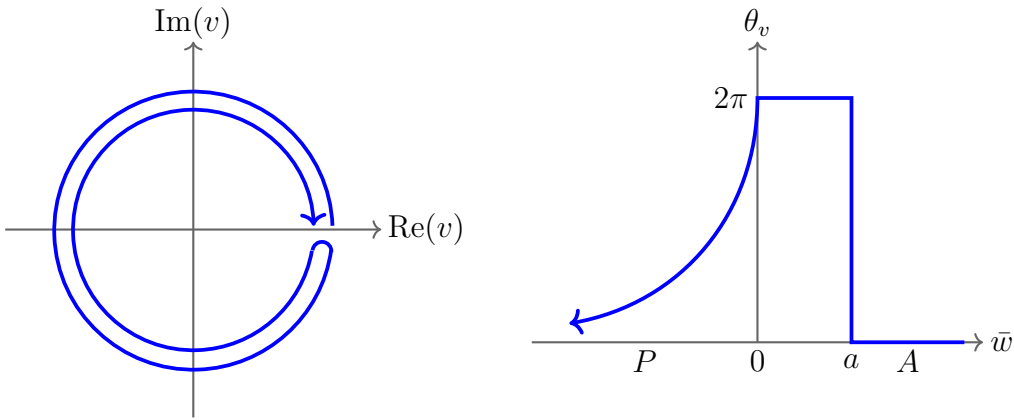
\begin{figure}[ht]
\begin{center}
\begin{tikzpicture}[dot/.style={draw,fill,circle,inner sep=1pt},scale=0.5]
%every node/.style={scale=0.8},decoration={markings,mark=at position 0.5 with {\arrow{>}}}]]
%
%
%\draw[very thick,fill=black!30,opacity=0.2] (0,0) circle (4cm);
%
%%
%\filldraw[black] (3.2,0) circle (0.1cm);
%\filldraw[black] (4.8,0) circle (0.1cm);
\draw[->,thick,black!60] (0,-5) -- (0,5);
\draw[->, thick,black!60] (-5,0) -- (5,0);
%\draw[->, very thick] (0,0) -- (2.61, 2.61 );
\node at (6.2,0) {{Re}$(v)$};
\node at (0,5.5) {{Im}$(v)$};
\draw[line width=0.5mm,blue] (3.7,0.1) arc (2:352:3.7cm);
\draw[<-, line width=0.5mm,blue] (3.2,0.1) arc (2:350:3.2cm);
\draw[line width=0.5mm,blue] (3.67,-0.6) arc (-5:175:0.25cm);
%\node at (-1.7,-4) {I};
%\node at (-1.1,-2.3) {II};
%
%
\begin{scope}[xshift=15cm,yshift=-3cm]
\draw[->,thick,black!60] (0,0) -- (0,8);
\draw[->, thick,black!60] (-6,0) -- (6,0);
\draw[blue, line width=0.5mm,->] (5.5,0) -- (2.5,0) -- (2.5,6.5) -- (0,6.5) to[out=-90,in=10] (-5,0.5);
\node at (2.5,-0.5) {$a$};
\node at (0,-0.5) {$0$};
\node at (4,-0.5) {$A$};
\node at (-3,-0.5) {$P$};
\node at (0,8.5) {$\theta_v$};
\node at (-0.7,6.5) {$2\pi$};
\node at (6.5,0) {$\bar w$};
\end{scope}
\end{tikzpicture}
\caption{\footnotesize  The anti-chiral operator exits the interval $A$ and the associated cross-ratio $v$ undergoes an anti-clockwise phase rotation taking it to a different Riemann sheet. Upon reaching the projection slit $\PR$, the rotation is reversed.  The argument of $v$ undergoes a step change from $0$ to 
$2\pi$ when the anti-chiral operator leaves $A$, and then relaxes to zero asymptotically after it enters $\PR$.}
\label{cross-ratio} 
\end{center}
\end{figure}

The first scenario we will be interested in is when $w\in A$. In this scenario, using \eqref{mapbeta} and \eqref{mapa}, the argument of the cross-ratio $u$ remains positive and close to zero when $w$ is deep inside the interval,
\EQ
{\theta_u\approx \begin{cases}\frac{4\pi \epsilon}{\beta}\, e^{-2\pi (w-a)/\beta}\qquad\qquad &\epsilon\ll\beta \ll a \ll w\\\\
2\sqrt\epsilon \,a\, w^{-3/2}\qquad & \epsilon\ll a\ll w\ll \beta\,,
\end{cases}
}
where we have shown both the high and low temperature asymptotics. In both cases
\EQ{
\frac{\sin\frac{\theta_u}{2}}{\sin \frac{\theta_u}{2n}}\approx n\ .
}
The phase of the cross-ratio $v$, constructed from the anti-chiral/left-moving coordinates has the same form (as a function of $\bar w$) when the anti-chiral coordinate $\bar w \in A$. However, whilst keeping $w$ fixed in $A$, the phase of  $v$  executes nontrivial motion as we take the anti-chiral insertions outside  $A$ past  the endpoint $\bar w =a$ to  eventually encounter the projection slit $\PR$ at $\bar w =0$.  As depicted in figure \eqref{cross-ratio}, $v$ first rotates anti-clockwise through a phase of $2\pi$ as the anti-chiral quench exits $A$,
\EQ
{%\theta_v\Big|_{\bar w\gg a}\approx \frac{4\pi \epsilon}{\beta} e^{-2\pi (\bar w-a)/\beta}\,,\qquad\qquad 
\theta_v\approx \begin{cases}2\pi -  4\pi \epsilon/\beta
\qquad\qquad &\epsilon\ll \beta\ll \bar w \ll a \\\\ 
2\pi -  2 \epsilon/\sqrt{a\bar w} &0 <\epsilon\ll\bar w < a\ll\beta
\end{cases}.
}
When it reaches $\PR$, i.e.~$\bar w<0$, the cross-ratio $v$ commences a clockwise rotation back towards  $v =1$ as $|\bar w|$ increases and the images of the operator insertion points $\bar \xi_1$ and $\bar \xi_2$ move away from each other along the boundary in the $\xi$-plane,
\EQ{
\theta_v\approx \begin{cases}4 a \,e^{-\pi |\bar w|/\beta} \qquad\qquad & |\bar w|\gg \beta\,,\quad \bar w<0\,,\\\\
4\sqrt{a/|w|} & \beta\gg |w|\gg a\,,\quad \bar w <0\,.
\end{cases}
}
Therefore, both when $\bar w\in A$ and when $\bar w\to-\infty$, we have $\theta_v\approx 0$ and 
\EQ
{\frac{\sin\frac{\theta_v}{2}}{\sin \frac{\theta_v}{2n}}\approx n\,,}
so the R\'enyi entropy \eqref{renyilargec}  is vanishingly small in these two regimes for all $n$. 

\begin{figure}[ht]
%\begin{tikzpicture} [scale=0.5,every node/.style={scale=0.8}]
%
%\begin{scope}[xshift=14cm,yshift=0cm]
%\node[right,fill=blue!20, draw=blue, thick, rounded corners=3pt] (a1) at (4,5) {$w\not\in A$};
%\draw[->,thick,black!60] (0,0) -- (0,8);
%\draw[->, thick,black!60] (-5,0) -- (6,0);
%\draw[blue, line width=0.5mm] (0,0) -- (-4,4);
%\draw[blue, line width=0.5mm] (2.7,6.5) -- (5.5,6.5);
%\draw[blue, line width=0.5mm] plot [smooth, tension=0.3] coordinates {(0,0)  (2.3,0.3) (2.7,6.5)};
%\draw[dotted,thick] (-2,6.5) -- (4,6.5);
%
%\node at (2.5,-0.5) {$a$};
%\node at (0,-0.5) {$0$};
%\node at (4,-0.5) {$A$};
%\node at (-2.5,-0.5) {$P$};
%\node at (0,8.5) {$S_n^{\ket{\cal O}}(A)$};
%\node at (6.5,0) {$\bar w$};
%\end{scope}
%
%\begin{scope}[xshift=0cm,yshift=0cm]
%\node[right,fill=blue!20, draw=blue, thick, rounded corners=3pt] (a1) at (4,5) {$w\in A$};
%\draw[->,thick,black!60] (0,0) -- (0,8);
%\draw[->, thick,black!60] (-5,0) -- (6,0);
%\draw[blue, line width=0.5mm] (-4.5,0) -- (-0.2,0);
%\draw[blue, line width=0.5mm] (2.7,0) -- (5.5,0);
%\draw[blue, line width=0.5mm] plot [smooth, tension=0.3] coordinates {(-0.2,0) (0.2,6) (1.25,6.4) (2.3,6) (2.7,0)};
%\draw[dotted,thick] (-4,6.5) -- (4,6.5);
%
%\node at (2.5,-0.5) {$a$};
%\node at (0,-0.5) {$0$};
%\node at (4,-0.5) {$A$};
%\node at (-2.5,-0.5) {$P$};
%\node at (0,8.5) {$S_n^{\ket{\cal O}}(A)$};
%\node at (6.5,0) {$\bar w$};
%\end{scope}
%\end{tikzpicture}
%\begin{figure}[ht]
\begin{center}
\includegraphics[width=2.5in]{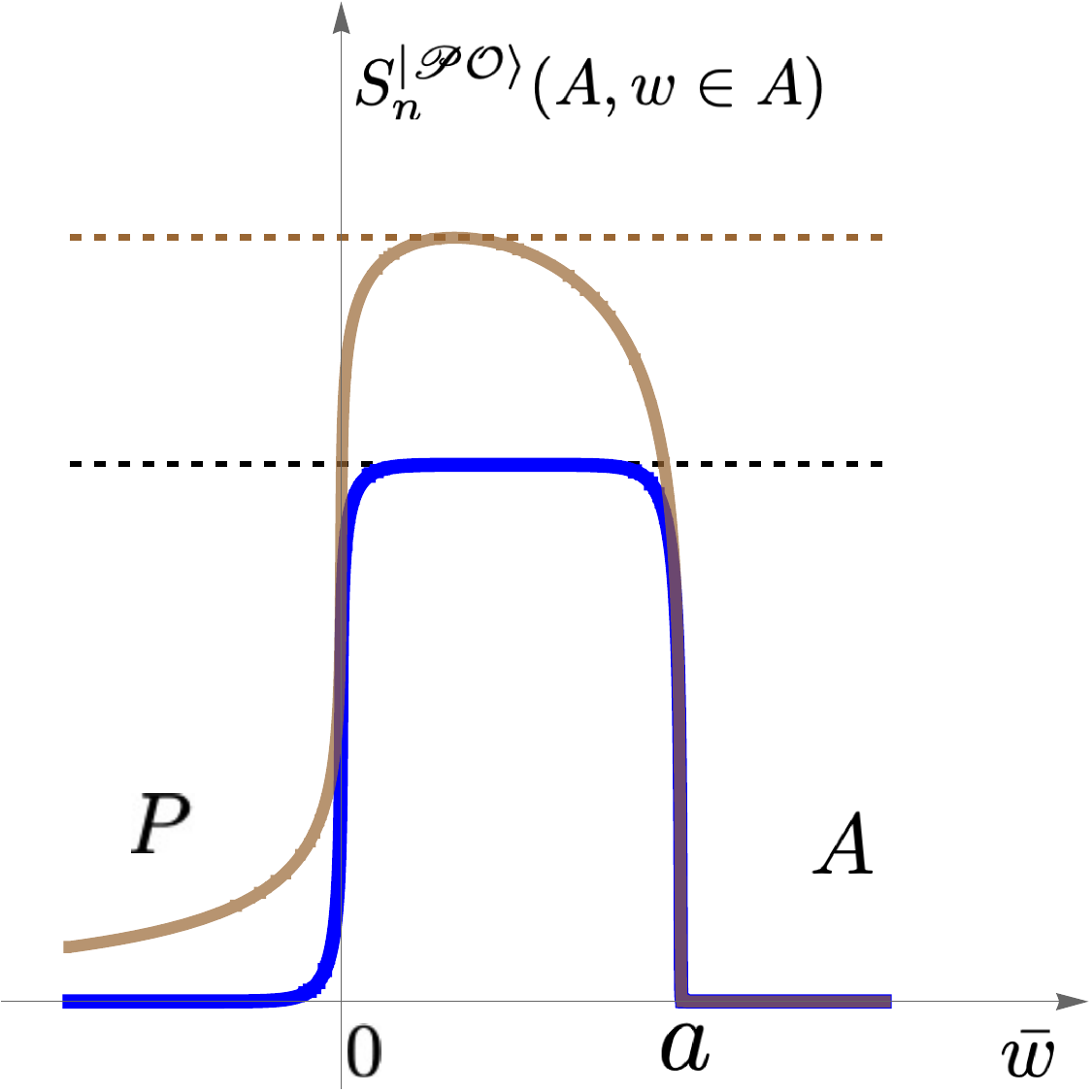} \hspace{1in}
\includegraphics[width=2.5in]{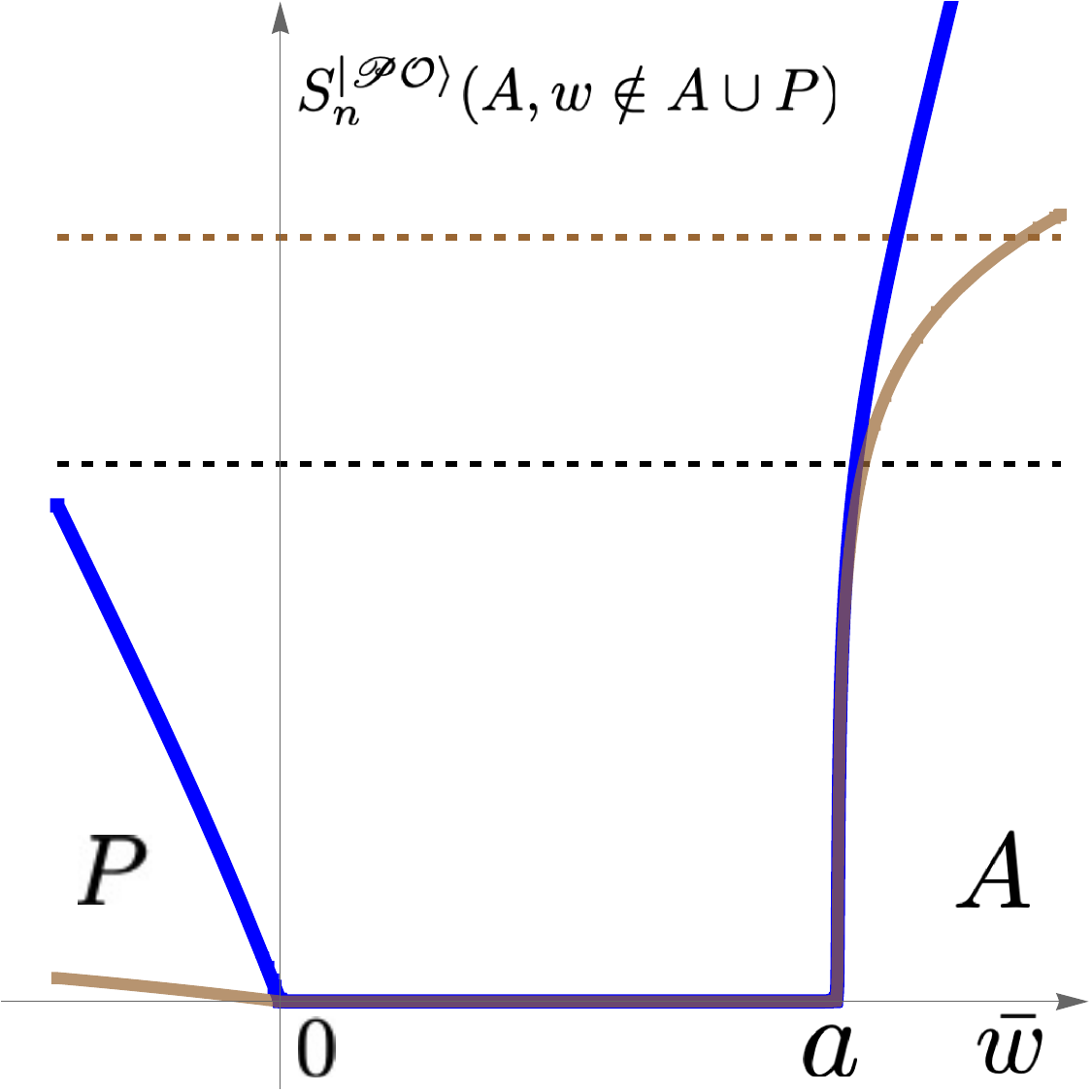}
%\hspace{1in}
%\includegraphics[width=2.0in]{renyi2.pdf}
\end{center}
\caption{\footnotesize Plots for R\'enyi entropy excess with $\epsilon=10^{-3}$, $a=1$, $n=5$, and two values of temperature, $\beta=0.3$ (blue) and $\beta=100\pi$ (brown). {\it Left:} Excess entropy as a function of $\bar w$ for $w\in A$, at zero temperature (brown) and large temperature (blue) . The peak values (dashed lines), when $\bar w\notin A\cup P$ in the large-$c$ theory, depend on the quench width parameter $\epsilon$, the temperature and $n$ as in eqs. \eqref{peak} and \eqref{peak0}. {\it Right:} The case when $w\notin A\cup P$ for the same parameters and vertical scale as figure on the left. As $\bar w$ leaves $A$, the R\'enyi entropy decreases to zero and when it enters the projection slit $\PR$, the entropy rises approximately linearly for the thermal theory (blue) and logarithmically at zero temperature (brown); the putative saturation of the curve  on either side is not visible in  the strict large-$c$ limit.}
\label{renyislit1}
\end{figure}

 At intermediate time scales when the anti-chiral operator lies between $A$ and $\PR$ i.e.~$\bar w\in[0,a]$, the R\'enyi entropy is non-vanishing, plateauing at high temperatures (and small widths $\epsilon\ll \beta$),
\EQ
{
S_n^{\ket{{\mathscr P}\cal O}}(A, w\in A,\bar w\notin A\cup P)= \frac{2n \Delta}{n-1} \log\left[\frac{\beta}{2\pi\epsilon} \,n \sin\frac{\pi}{n}\right]\, \qquad \qquad \beta\ll \bar w < a\,. \label{peak} 
}
In the zero temperature theory, the entropy peaks inside the interval $[0,a]$
at $\bar w = a/3$ for a small width quench ($\epsilon\ll a$), with  peak value,
\EQ
{
S_n^{\ket{{\mathscr P}\cal O}}(A, w\in A,\bar w\notin A\cup P)= \frac{2n \Delta}{n-1} \log\left[\frac{2a}{3\sqrt 3\epsilon } \,n \sin\frac{\pi}{n}\right]\, \qquad \quad \bar w=\tfrac13a\,,\quad \beta\gg a\,. \label{peak0} 
}
The behaviour of the entropy as a function of $\bar w$ is shown in figure \ref{renyislit1}. We pause here to note that the plateau value is not given by the log of the quantum dimension, and is in fact non-universal since it depends on the pulse width $\epsilon$ and the temperature. In the large-$c$ limit, the expectation is that the quantum dimension diverges nonperturbatively with $c$. On the other had, at finite temperature and for a finite width pulse we expect some of the entanglement to be carried out by the spread of the excitation and/or thermal fluctuations. The lowering of the saturation value by thermal fluctuations at finite width is also seen in  the free boson and Ising models \cite{Caputa:2014eta}.

The large-$c$ result is qualitatively distinct from what we have seen for free scalar CFT in \eqref{pwina} where except for $n=2$, we see the R\'enyi excess not change (odd $n$) or only decrease by a small amount (even $n$), generically  non-vanishing when the anti-chiral mode is deep in the projection slit (all the while $w\in A$).

\begin{figure}[ht]
\begin{center}
\includegraphics[width=2.5in]{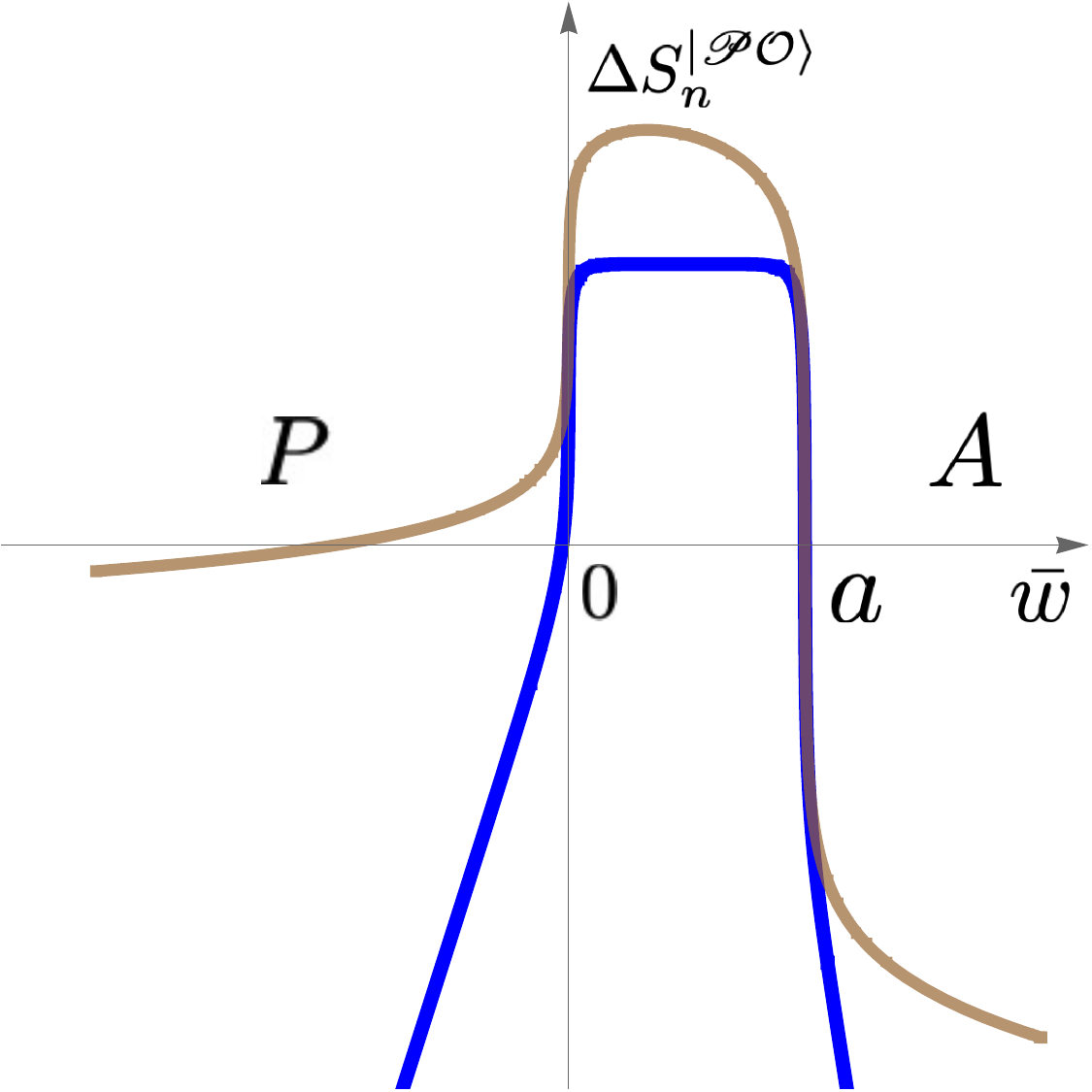} 
\end{center}
\caption{\footnotesize The entropy difference $\Delta S_n^{\ket{{\mathscr P}{\cal O}}}(\bar w)$ defined in  for the large-$c$ theory, plotted for low (brown) and high (blue) temperatures. Negative value of $\Delta S_n^{\ket{{\mathscr P}{\cal O}}}$ when the antichiral position is in $P$ shows successful teleportation.}
\label{difflargec}
\end{figure}

\paragraph{Scenario II $w\notin A$:} In this case, when the anti-chiral point 
$\bar w\in[0,a]$, i.e. not in $A\cup P$, we expect vanishing excess entropy for $A$. Therefore, we choose the branch of ${\cal C}_n(u,v)$ in \eqref{hhllll} with vanishingly small  $\theta_v$ and $\theta_u$ in this state:
\EQ
{
\theta_v=\theta_u\big|_{w,\bar w\notin A, \PR } \approx \begin{cases}-  4\pi \epsilon/\beta
\qquad\qquad &\epsilon\ll\beta\ll \bar w  \\\\ 
 -  2 \epsilon/\sqrt{a\bar w} & 0  < \epsilon\ll\bar w < a\ll\beta
\end{cases}.
}
If $\bar w\in A$, then the excess entropy picks up the entropy of the operator quench, as expected from the analysis without a projection. On the other hand, moving $\bar w$ into $\PR$ then results in $\theta_v$ undergoing a clockwise rotation to 
$\theta_v\approx-2\pi$,
\EQ
{
\theta_v\big|_{\bar w\in P}\approx \begin{cases}-2\pi + 4 a \,e^{-\pi |\bar w|/\beta} \qquad\qquad & |\bar w|\gg \beta\,,\\\\
-2\pi + 4\sqrt{a/|\bar w|} & \beta\gg |\bar w|\gg a\,.
\end{cases}
}
For large $|\bar w|$ in $\PR$ the excess R\'enyi entropy grows monotonically,
\EQ
{
S_n^{\ket{{\mathscr P}\cal O}}(A, w\notin A\cup P,\bar w\in P)\to\begin{cases}  \frac{2 n\Delta}{n-1} \,\frac{\pi}{\beta}\, |\bar w|\qquad &\beta\ll a \ll |\bar w|\\\\
\frac{2 n\Delta}{n-1} \,\log\left[\tfrac{n}{2}\sin\frac{\pi}{n}\sqrt{\frac{|\bar w|}{a}}\right]\qquad &\beta\gg |\bar w|\gg a\,. 
\end{cases}
}
This is consistent with the expectation that as $\bar w$ moves into $\PR$, the state of the anti-chiral excitation is teleported to $A$ so the entanglement of $A$ rises. Finally, we plot the UV safe entropy difference $\Delta S_n^{\ket{{\mathscr P}{\cal O}}}(\bar w)$ defined in \eqref{fet}  for the large-$c$ theory in figure \ref{difflargec}. The asymptopia for this quantity follow immediately from the analysis above.

\section{Finite regions}\label{s5}

In this section we consider what happens to the entropy when the projection region $\PR$ and entropy region $A$ are both finite. With finite regions, we are only able to uniformize the replica geometry when $n=2$, so our analysis will only allow us to compute the second R\'enyi entropy. 

\begin{figure}[ht]
\begin{center}
\begin{tikzpicture} [scale=0.8,every node/.style={scale=0.8}]
\draw[fill = Plum!10!white, draw = Plum!10!white] (0,-4) rectangle (6,4);
\draw[fill = yellow!10!white, draw = yellow!10!white] (0,-4) rectangle (-6,4);
\draw[black!40,thick,] (-6,0) --(6,0);
\draw[black!40,thick,] (-6,4) --(6,4);
\draw[black!40,thick,] (-6,-4) --(6,-4);
\draw[red,thick] (0,0) -- (0,4);
\draw[green,thick] (0,0) -- (0,-4);
\draw[red,thick,dashed] (-6,-4) -- (-6,0);
\draw[red,thick,dashed] (6,-4) -- (6,0);
\draw[green,thick,dashed] (-6,0) -- (-6,4);
\draw[green,thick,dashed] (6,0) -- (6,4);
\draw[blue,thick] (3,-4) -- (3,4);
\draw[blue,thick,dashed] (-3,-4) -- (-3,4);
%
%\begin{scope}[xshift=-0.9cm]
%\node at (6.5,3.5) {$z$};
%\draw[-] (6.2,3.9) -- (6.2,3.2) -- (6.8,3.2);
%\end{scope}
%
%
\filldraw[black] (2.9,2.5) circle (2pt);
\filldraw[black] (3.1,2.5) circle (2pt);
\filldraw[black] (-0.2,2.5) circle (2pt);
\filldraw[black] (-0.2,-2.5) circle (2pt);
\filldraw[black] (2.9,-2.5) circle (2pt);
\filldraw[black] (3.1,-2.5) circle (2pt);
\filldraw[black] (-5.8,2.5) circle (2pt);
\filldraw[black] (-5.8,-2.5) circle (2pt);
\filldraw[black] (-0.15,0.15) circle (2pt);
\filldraw[black] (-0.15,-0.15) circle (2pt);
\filldraw[black] (-5.8,0.15) circle (2pt);
\filldraw[black] (-5.8,-0.15) circle (2pt);
\filldraw[purple] (2,0.1) circle (2pt);
\filldraw[purple] (2,-0.1) circle (2pt);
\filldraw[purple] (4,0.1) circle (2pt);
\filldraw[purple] (4,-0.1) circle (2pt);
\node at (2,0.5) {$z_1$};
\node at  (2,-0.5) {$z_2$};
\node at (4,0.5) {$z_4$};
\node at  (4,-0.5) {$z_3$};
\node at (2.4,2.5) {$z_1$};
\node at  (2.4,-2.5) {$z_2$};
\node at (-0.6,2.5) {$\bar z_2$};
\node at  (-0.6,-2.5) {$\bar z_1$};
\node at (3.6,2.5) {$z_4$};
\node at  (3.6,-2.5) {$z_3$};
\node at (-5.4,2.5) {$\bar z_3$};
\node at  (-5.4,-2.5) {$\bar z_4$};
\draw[->,thick,densely dashed] (-0.15,0.15) -- (-0.15,2.3);
\draw[->,thick,densely dashed] (-0.15,-0.15) -- (-0.15,-2.3);
\begin{scope}[rotate=180,xshift=6cm]
\draw[->,thick,densely dashed] (-0.15,0.15) -- (-0.15,2.3);
\draw[->,thick,densely dashed] (-0.15,-0.15) -- (-0.15,-2.3);
\end{scope}
%
%\node at (-4.5,3.6) {bottom};
%\node at (-1.5,3.6) {top};
%\node at (-4.5,3) {image};
%\node at (-1.5,3) {image};
%
%\node at (1.5,3.6) {top};
%\node at (4.5,3.6) {bottom};
%
\node at (0,4.3) {$iK'$};
\node at (0,-4.3) {$-iK'$};
\node at (-6.5,0) {$-2K$};
\node at (6.4,0) {$2K$};
%
%\node at (1.5,4) {$C$};
%\node at (1.5,-4) {$C$};
\node at (0.2,1.5) {$\PR$};
\node at (0.2,-1.5) {$\PR$};
%\node at (1,0) {$B$};
\node at (2.8,1.5) {$A$};
\node at (2.8,-1.5) {$A$};
\end{tikzpicture}
\caption{\footnotesize The uniformatization required to compute $\Delta S_2^{\ket{\cal O}}(A)$ with finite regions $\PR$ and $A$.}
\label{fig7} 
\end{center}
\end{figure}
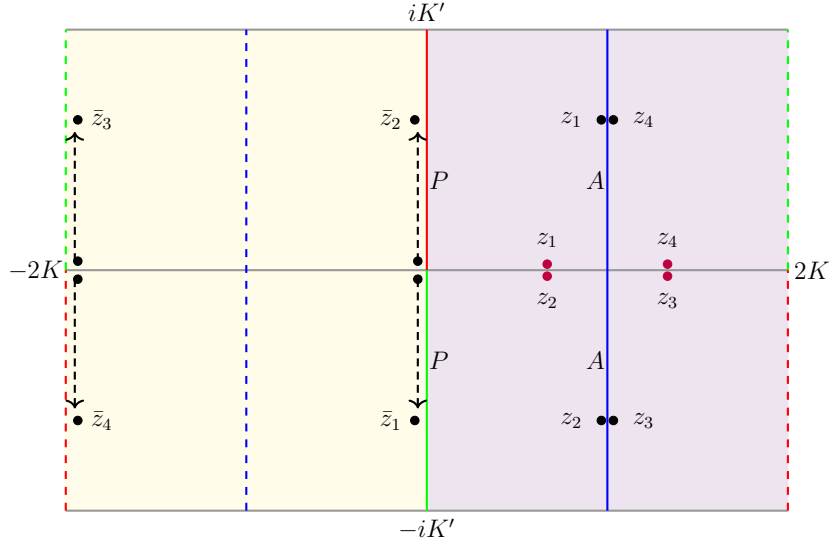

Let us take $\PR=[-p,0]$ and $A=[a,b]$. The map to the BCFT picture is now via
\EQ{
\xi=\sqrt{\frac{b+p}b}\sqrt{\frac w{w+p}}\ .
}
The entropy region $A$ is mapped to $[k,1]$, where we define the cross ratio
\EQ{
k=\sqrt{\frac{(b+p)a}{(a+p)b}}\ .
}
For the case $n=2$, the replica geometry is a torus and we can uniformize with an appropriate map involving  a Jacobi elliptic function
\EQ{
\xi=k\,\text{sn}(z,k^2)\ ,
}
where $K=K(k)$ is the complete elliptic integral of the first kind and $k$ plays the r\^ole of the elliptic modulus. The uniformized geometry is a the rectangle defined as the $z$ plane modulo integer combinations of $4K$ and $2iK'$. We take a fundamental region $[-2K,2K]\times[-iK',iK']$, as shown in figure \ref{fig7}.

For our scalar field theory, the 4-point function on the torus takes the same \eqref{gug} but with the replacement
\EQ{
z_{jk}\longrightarrow \theta_1(u_{jk},q)e^{-\IM(u_{jk})^2/(\pi\IM\tau)}\ ,
}
where the re-scaled $u=\pi z/4K$ fundamental region is defined by periodicities $\pi$ and $\pi\tau=\pi K'/2K$ and $q=e^{i\pi\tau}$. Note that the elliptic function above is doubly-periodic modulo a phase. The cross ratio of the anti-chiral points \eqref{piw} is a doubly-periodic function. Rather than work in terms of theta functions, we find it more convenient tor§1 translate in terms of Jacobi elliptic functions. The anti-chiral  points are all determined in terms of, say, $\bar z_1$:
\EQ{
\bar z_2=-\bar z_1\ ,\qquad \bar z_3=-\bar z_1-2K\ ,\qquad \bar z_4=\bar z_1-2K\ ,
}
the cross ratio $\bar\chi$, as a function of $\bar z_1$, has double zeros at $\bar z_1=2n_1K+in_2K'$, $n_i\in{\mathbb Z}$ and double poles at $\bar z_1=(2n_1+1)K+in_2K'$. Matching the behaviour at the zeros and poles uniquely fixes
\EQ{
\bar\chi=-\frac{\SN^2(\bar z_1,k^2)}{(1-\SN^2(\bar z_1,k^2))(1-k^2\SN^2(\bar z_1,k^2))}\ .
}
On the other hand
\EQ{
\SN(z,k^2)=i\sqrt{\frac{(a+p)|\bar w|}{(p-|\bar w|)a}}\ ,
}
which gives
\EQ{
\bar\chi=\frac{b(a+p)(p-|\bar w|)|\bar w|}{p^2(a+|\bar w|)(b+|\bar w|)}\ .
}
When $\bar w=0$ or $\bar w=-p$, we have $\bar\chi=0$ and so the channel $\Be$ dominates as expected because the anti-chiral point is only on the edge of $\PR$ and the entropy shift is $\Delta S_2^{\ket{\cal O}}(\bar w=0)=\log d_{\cal O}$. As $\bar w$ enters $P$, $\bar\chi$ rises to a maximum equal to $d_{\cal O}(1+k)^{-2}$. At this point the entropy shift $\Delta S_2^{\ket{\cal O}}(\bar w)$ is a minimum. Figure \ref{fig8} shows the behaviour of $\Delta S_2^{\ket{\cal O}}(\bar w)$ as $|\bar w|$ varies from $0$ to $p$. 
\pgfdeclareimage[interpolate=true,width=9cm]{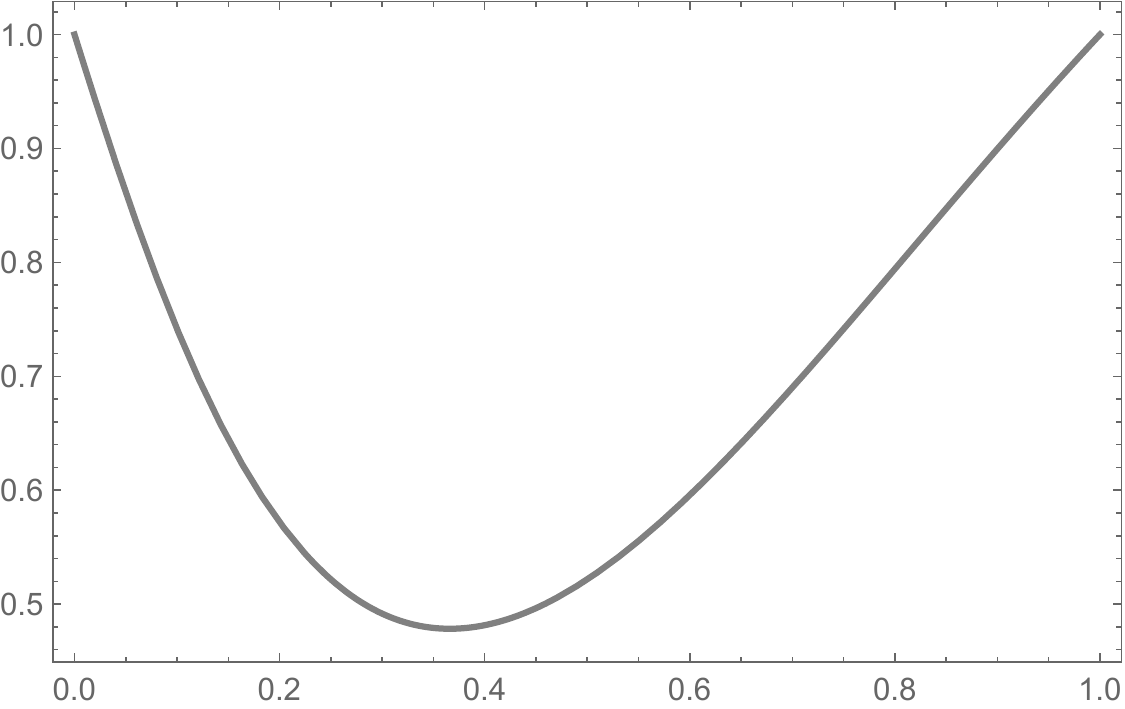}{pic5}
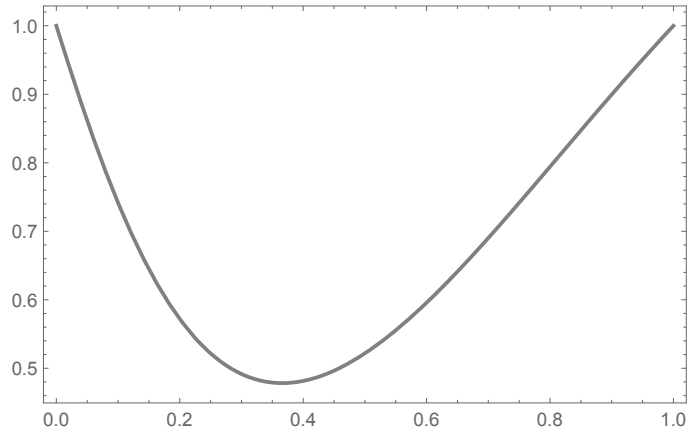
\begin{figure}
\begin{center}
\begin{tikzpicture}[scale=1,every node/.style={scale=0.7}]
\pgftext[at=\pgfpoint{0cm}{0cm},left,base]{\pgfuseimage{pic5}} 
\end{tikzpicture}
\caption{\footnotesize The entropy shift $\Delta S_2^{\ket{\cal O}}(\bar w)$ in units of $\log d_{\cal O}$ as a function of $|\bar w|$ for $\bar w\in[-p,0]$ for $p=1$, $a=1$ and $b=2$.}
\label{fig8}
\end{center}
\end{figure}

\section{Discussion}

We have set up a scenario to investigate teleportation in a CFT using the fact that an inner product of states with a local operator quench and with a projector onto a Cardy state in a subregion of a Cauchy surface can be mapped to a correlator in BCFT. The motivation was to shed light into how information actually emerges from a black hole using new insights \cite{Akers:2022qdl} related to the final state projection hypothesis \cite{Horowitz:2003he}. In the black hole setting, the projector that acts on the Hilbert space of the effective modes inside the black hole is automatically tuned to ensure perfect teleportation and no loss of information. In our analysis the projector is fixed by the Cardy state and we have shown that with the CFT in the vacuum state the R\'enyi entropies point to a less than 100\% efficiency of teleportation for free scalar fields. For large-$c$ sparse CFTs the results point toward more efficient teleportation, but are incomplete as the saturated values of the R\'enyi entropies are not accessible in this limit.

The results (section \ref{s3.2}) for the R\'enyi excesses in the free scalar CFT are surprising and even counterintuitive. The difference between even and odd $n$ is pronounced and follows robustly from universal combinatorial factors associated to the OPE limits of replica configurations for local quench operators as they are placed deep in the projection region. A closely related puzzle is the resulting ambiguity in the von Neumann limit $n\to 1$. Extending this study to cover a wider class of CFTs and potentially tractable example cases is  clearly desirable. The large-$c$ analysis on the other hand, is somewhat incomplete due to known large-$c$ effects, related to  putative/conjectured divergence of the quantum dimension  in the large-$c$ limit. This precludes a clear conclusion  on the efficacy of teleportation.

Finally, it is important to stress that we have  explored  the entanglement structure of a one parameter family of initial states prepared with (entangled) chiral-antichiral insertions in the different subregions within the CFT. We have not pursued the altogether more difficult problem of real time evolution of the system prepared in a local quench state subjected to a projection (as the projection does not commute with the Hamiltonian).  This is clearly the more interesting dynamical question, also deeply relevant for the black hole evaporation problem.

\acknowledgments{TJH and SPK would like to acknowledge support from STFC Consolidated Grant Award ST/X000648/1. ZG is supported by an  STFC PhD studentship award. }
\vspace{1cm}

{\footnotesize Open Access Statement - For the purpose of open access, the authors have applied a Creative Commons Attribution (CC BY) licence to any Author Accepted Manuscript version arising. 

Data access statement: no new data were generated for this work.}

\end{document}